\documentclass{article}
\usepackage{emulateapj,apjfonts}
\submitted{Received: 2004 December 20, accepted: 2005 May 9, ApJ 631, 280} 
\input psfig

\def\ts{\thinspace}
\def\gapprox{$_>\atop{^\sim}$} 
\def\lapprox{$_<\atop{^\sim}$}

\newdimen\sa  \def\sd{\sa=.1em \ifmmode $\rlap{.}$''$\kern -\sa$
                               \else \rlap{.}$''$\kern -\sa\fi}

\begin{document}

\lefthead{The Triple Nucleus and Supermassive Black Hole of M{\thinspace}31}

\righthead{Bender et al.}

\title{HST STIS Spectroscopy of the Triple Nucleus of M{\thinspace}31:\\Two
Nested Disks in Keplerian Rotation around a Supermassive Black Hole}

\author{
Ralf Bender\altaffilmark{1,2,3},  
John Kormendy\altaffilmark{4}, 
Gary Bower\altaffilmark{5}, 
Richard Green\altaffilmark{6}, 
Jens Thomas\altaffilmark{1,2},  
Anthony C.~Danks\altaffilmark{7},
Theodore Gull\altaffilmark{8}, 
J.~B.~Hutchings\altaffilmark{9}, 
C.~L.~Joseph\altaffilmark{10}, 
M.~E.~Kaiser\altaffilmark{11}, 
Tod R. Lauer\altaffilmark{6}, 
Charles H. Nelson\altaffilmark{12}, 
Douglas Richstone\altaffilmark{13},
Donna Weistrop\altaffilmark{12}, and
Bruce Woodgate\altaffilmark{8}
}

\altaffiltext{1}{Universit\"ats-Sternwarte, Scheinerstrasse 1,
M\"unchen 81679, Germany; bender@usm.uni-muenchen.de, jthomas@usm.uni-muenchen.de}

\altaffiltext{2}{Max-Planck-Institut f\"ur Extraterrestrische Physik,
Giessenbachstrasse, 85748 Garching-bei-M\"unchen, Germany; 
bender@mpe.mpg.de}

\altaffiltext{3}{Beatrice M. Tinsley Centennial Visiting Professor, 
University of Texas at Austin}

\altaffiltext{4}{Department of Astronomy, University of Texas, Austin,
Texas 78712; kormendy@astro.as.utexas.edu
}

\altaffiltext{5}{Computer Sciences Corporation, Space Telescope Science
Institute, 3700 San Martin Dr., Baltimore, MD 21218; bower@stsci.edu}

\altaffiltext{6}{National Optical Astronomy Observatories, P. O. Box
26732, Tucson, AZ 85726; green@noao.edu, lauer@noao.edu
}

\altaffiltext{7}{Emergent-IT, 1315 Peachtree Court, Bowie, MD 20721;
danks@yahoo.com}

\altaffiltext{8}{NASA/Goddard Space Flight Center, Code 681,
Greenbelt, MD 20771; gull@sea.gsfc.nasa.gov; woodgate@stars.gsfc.nasa.gov}

\altaffiltext{9}{Herzberg Institute of Astrophysics, National Research Council
                 of Canada, Victoria V9E 2E7, Canada; john.hutchings@hia.nrc.ca}

\altaffiltext{10}{Dept.~of Physics \& Astronomy, Rutgers University,
P.~O.~Box 849, Piscataway, NJ 08855; cjoseph@physics.rutgers.edu}
 
\altaffiltext{11}{Department of Physics \& Astronomy, Johns Hopkins
University, Homewood Campus, Baltimore, MD 21218; kaiser@pha.jhu.edu}

\altaffiltext{12}{Department of Physics, University of Nevada, 4505
S.~Maryland Parkway, Las Vegas, NV 89154; cnelson@physics.unlv.edu,
weistrop@physics.unlv.edu}

\altaffiltext{13}{Dept. of Astronomy, Dennison Bldg., Univ. of
Michigan, Ann Arbor 48109; dor@umich.edu}
 

\pretolerance=15000  \tolerance=15000

\begin{abstract} 
We present {\it Hubble Space Telescope\/} (HST) spectroscopy of the
nucleus of M{\thinspace}31 obtained with the Space Telescope Imaging
Spectrograph (STIS).  Spectra that include the Ca II infrared triplet
($\lambda \simeq 8500$ \AA)~see only the red giant stars in the double
brightness peaks P1 and P2.  In contrast, spectra taken at $\lambda\lambda
\simeq$ 3600{\ts}--{\ts}5100\ts\AA\ are sensitive to the tiny blue nucleus
embedded in P2, the lower-surface-brightness nucleus of the galaxy.  
P2 has a K-type spectrum, but we find that the blue nucleus has an A-type 
spectrum -- it shows strong Balmer absorption lines.  Hence, the blue 
nucleus is not blue because of AGN light but rather because it is 
dominated by hot stars.  We show that the spectrum is well described by 
A0 giant stars, A0 dwarf stars, or a 200-Myr-old, single-burst stellar 
population.  White dwarfs, in contrast, cannot fit the blue nucleus 
spectrum.  Given the small likelihood for stellar collisions, recent 
star formation appears to be the most plausible origin of the blue nucleus.
In stellar population, size, and velocity dispersion, the blue nucleus is
so different from P1 and P2 that we call it P3 and refer to the nucleus 
of M{\ts}31 as triple.

      Because P2 and P3 have very different spectra, we can make a
clean decomposition of the red and blue stars and hence measure the
light distribution and kinematics of each uncontaminated by the other.
The line-of-sight velocity distributions of the red stars near P2
strengthen the support for Tremaine's (1995) eccentric disk
model. Their wings indicate the presence of stars with velocities of
up to $~1000$ km s$^{-1}$ on the anti-P1 side of P2.

      The kinematic properties of P3 are consistent with a circular stellar
disk in Keplerian rotation around a supermassive black hole. If the P3
disk is perfectly thin, then the inclination angle $i \simeq 55^\circ$
is identical within the errors to the inclination of the eccentric disk
models for P1 $+$ P2 by Peiris \& Tremaine (2003) and by Salow \& 
Statler (2004).  Both
disks rotate in the same sense and are almost coplanar.  The observed
velocity dispersion of P3 is largely caused by blurred rotation and
has a maximum value of $\sigma = 1183 \pm 201$ km s$^{-1}$.  This is
much larger than the dispersion $\sigma \simeq 250$ km s$^{-1}$ of the
red stars along the same line of sight and is the largest integrated
velocity dispersion observed in any galaxy.  The rotation curve of P3
is symmetric around its center.  It reaches an observed velocity of $V
= 618 \pm 81$ km s$^{-1}$ at radius 0\farcs05 = 0.19 pc, where the
observed velocity dispersion is $\sigma = 674 \pm 95$ km s$^{-1}$.
The corresponding circular rotation velocity at this radius is $\sim$
1700 km s$^{-1}$.  We therefore confirm earlier suggestions that the
central dark object interpreted as a supermassive black hole is
located in P3.

      Thin disk and Schwarzschild models with intrinsic axial ratios 
$b/a$ \lapprox \ts0.26 corresponding to inclinations between $55^\circ$ 
and $58^\circ$ match the P3 observations very well. Among these models, 
the best fit and the lowest black hole mass are obtained for a thin disk 
model with $M_\bullet = 1.4\times 10^8\,M_\odot$.  Allowing P3 to have 
some intrinsic thickness and considering possible systematic errors, the
1-$\sigma$ confidence range becomes (1.1 to 2.3)\ts$\times 10^8\, M_\odot$.
The black hole mass determined from P3 is independent of but consistent 
with Peiris \& Tremaine's mass estimate based on the eccentric disk model 
for P1~$+$~P2.  It is $\sim$\ts2 times larger than the prediction 
by the correlation between $M_\bullet$ and bulge velocity dispersion 
$\sigma_{\rm bulge}$.  Taken together with other reliable black hole mass
determinations in nearby galaxies, notably the Milky Way and M{\ts}32,
this strengthens the evidence that the $M_\bullet$ -- $\sigma_{\rm
bulge}$ relation has significant intrinsic scatter, at least at low
black hole masses.

      We show that any dark star cluster alternative to a black hole
must have a half-mass radius \lapprox \ts$0\farcs03 = 0.11$ pc
in order to match the observations.  Based on this, M{\thinspace}31
becomes the third galaxy (after NGC 4258 and our Galaxy) in which
clusters of brown dwarf stars or dead stars can be excluded on
astrophysical grounds. 

\end{abstract}

\section{Introduction}

\pretolerance=15000 \tolerance=15000

      M{\thinspace}31 was the second\footnote{The first was
M{\thinspace}32 (Tonry 1984, 1987).  In retrospect, the resolution was
barely good enough for a successful BH detection (Kormendy 2004); 
i.{\thinspace}e., the BH was discovered essentially as early as possible.}
{\kern -1.5pt}galaxy in which stellar dynamics revealed the presence of 
a supermassive black hole (BH) (Kormendy 1987, 1988; Dressler \& Richstone
1988).  The spatial resolution of the discovery spectra was FWHM 
$\sim$\ts1\arcsec.  Axisymmetric dynamical models implied BH masses of
\hbox{$M_\bullet = (1~\rm to~10)\times 10^7$ $M_\odot$.} The smallest 
masses were given by disk models and the largest were given by spherical
models.  

      In 1988, it was already known that axisymmetry is only an approximation
to a more complicated structure. With {\it Stratoscope II}, Light et al.~(1974)
had observed that the nucleus is asymmetric.  The brightest point is offset 
both from the center of the bulge (Nieto et al.~1986) and from the velocity
dispersion peak (Dressler 1984; Dressler \& Richstone 1988; Kormendy (1988).
Then, using HST, Lauer et al.~(1993) discovered that the nucleus is double.
The brighter nucleus, P1, is offset from the bulge center by $\sim 0\farcs5$.
The fainter nucleus, P2, is approximately at the bulge center.  Early concerns
that an apparently double structure might only be due to dust were laid to 
rest when infrared images proved consistent with optical and ultraviolet images
(Mould et al.~1989; Rich et al.~1996; Davidge et al.~1997; Corbin, O'Neil, \&
Rieke 2001).  These results were confirmed at higher resolution and
signal-to-noise using WFPC2 (Lauer et al.~1998).  With the discovery of the
double nucleus, work on the central parts of M{\thinspace}31 went into high
gear.

      Bacon et al.~(1994, 2001) used integral-field spectroscopy to map
the two-dimensional velocity field near the center of M{\thinspace}31.  
They found that the kinematical major axis of the nucleus is not the 
same as the line that joins P1 and P2.  The rotation curve is
approximately symmetric about P2, i.{\thinspace}e., about the center of 
the bulge.  However, this is not the point of maximum dispersion.  
Instead, the brightest and hottest points are displaced from the rotation
center by similar amounts in opposite directions. 

      The above results created two acute needs.  First, the rich 
phenomenology of the double nucleus cried out for explanation.  Second, 
the P1\thinspace--\thinspace{P2} asymmetry raised doubts about BH 
mass measurements.  This paper is mainly about the BH.  HST allows 
us to take an important step inward by studying a blue cluster of stars 
embedded in P2.  We introduce this cluster in \S\thinspace1.1.  Second, 
our spectroscopy of P1 $+$ P2 (\S\S\thinspace2 and 3) provides further 
support for the preferred model of the double nucleus (Appendix).  
Since that model affects much of our discussion, we summarize it 
in \S\ts1.2.  For comprehensive reviews, see Peiris \& Tremaine (2003)
and Salow \& Statler (2004). 

\subsection{P3: The Blue Star Cluster Embedded in P2}

      Nieto et al.~(1986), using a photon-counting detector on the
Canada-France-Hawaii Telescope (CFHT), were the first to illustrate that P2 is
brighter than P1 at 3750{\ts}\AA~(contrast their Figure 3 with Figure 4 in
Light et al.~1974; cf.~Figure~3 here). However, they did not realize
this.~Instead, they focused on the strong color gradient -- bluer inward
-- and worried because this was inconsistent with published data.  But 
these data were taken in the red or else had poor spatial resolution; they 
could not have seen the ultraviolet center.  Nieto and collaborators found no 
problem with their data but concluded that ``Further observations
are required to settle this question.''

      King et al.~(1992) confirmed the ultraviolet excess in the nucleus
using the HST Faint Object Camera (FOC) at 1750\ts\AA. Using the same image,
Crane et al.~(1993b) illustrated that P2 is brighter than P1 but did not 
comment on this.  Bertola et al.~(1995) illustrated the same effect using 
FOC $+$ F150W $+$ F130LP images but again did not comment that it is P2, 
not P1, that is brighter in the ultraviolet.

      Therefore, it was King, Stanford \& Crane (1995) who discovered that P2
is much brighter than P1 in the ultraviolet.  This result was again based on 
the 1750 \AA~FOC images.  The blue light comes from a compact source that is 
embedded in P2 and that is similar in color and brightness to
post-asymptotic-giant-branch (PAGB) stars seen elsewhere in the bulge
(King et al.~1992; Bertola et al.~1995).  King et al.~(1995) proposed that 
the source might be nonthermal light from the weak AGN that is detected in 
the radio (Crane et al.~1992, 1993a), although they recognized that it could 
be a single PAGB star.  Subsequently, Lauer et al.~(1998) and Brown et 
al.~(1998) resolved the source; its half-power radius is $\simeq$ 
0\farcs06 = 0.2 pc.  Both papers argued that it is a cluster of stars.  
Lauer et al.~(1998) combined the King et al.~(1995) UV fluxes with optical 
fluxes to conclude that the source is consistent with an A-star spectrum.  

      In this paper we present STIS spectra and show directly that the 
source is composed of A stars (\S\ts4).  We also demonstrate that it is most
consistent with a disk structure rather than with a dynamically hot cluster 
(\S\S\ts5, 6).  Because the blue cluster is so distinct from P1 $+$ P2 in 
terms of stellar content and kinematics, we call it P3 and refer to the 
nucleus of M{\thinspace}31 as triple.

      The disk structure of P3 allows us to make a new and more reliable
measurement of the central dark mass (\S\ts7). From the kinematics of P3 we also 
show that the dark object must be confined inside a radius $r$ \lapprox
\ts0\farcs03 = 0.11 pc.  This implies that alternatives to a BH, such as a
cluster of brown dwarf stars or stellar remnants, are inconsistent with the
observations (\S\ts8).

\subsection{The Eccentric Disk Model of P1 $+$ P2}

       Tremaine (1995) proposed what is now the standard model of P1 and P2.
His motivation was the realization (see also Emsellem \& Combes 1997) that
the simplest alternative -- an almost-completed merger -- is implausible.
Two clusters in orbit around each other at a projected separation of 
$0\farcs49 = 1.8$~pc would merge in \lapprox $10^8$ yr by dynamical friction.
Instead, Tremaine proposed that both nuclei are parts of the same eccentric 
disk of stars.  The brighter nucleus, P1, is farther from the BH and
results from the lingering of stars near apocenter.  The fainter nucleus, 
P2, is explained by increasing the disk density toward the center.  A BH 
is required in P2 to make the potential almost Keplerian; only then might 
the alignment of orbits be maintained by the modest self-gravity of the disk.

     Statler et al.~(1999), Kormendy \& Bender (1999, hereafter KB), and 
Bacon et al.~(2001) showed that the nucleus has the signature of the 
eccentric disk model.  The most direct evidence is the asymmetry in 
$V(r)$ and $\sigma(r)$.  Eccentric disk stars should linger at apocenter 
in P1; $V$~and~$\sigma$ are observed to be relatively small there.  The
same stars should pass pericenter in P2, slightly on the anti-P1 side of
the BH; the velocity amplitude is observed to be high on the anti-P1
side of the blue cluster.  Because the PSF and the slit blur light from
stars seen at different radii and viewing geometries, the apparent
velocity dispersion should also have a sharp peak slightly on the anti-P1
side of the BH.  All of the above papers demonstrated that the dispersion 
has a sharp peak in P2.  KB showed further that the $\sigma$ peak is 
slightly on the anti-P1 side~of~the~blue~cluster.  Therefore, they 
suggested that the BH is in the blue~cluster.  Finally, KB demonstrated
that the spectra and metal line strengths of P1 and P2 are similar to 
each other but different from those of the bulge.  
Therefore P1 cannot be an accreted globular cluster or dwarf galaxy.

      Peiris \& Tremaine (2003) refined the eccentric disk model to 
optimize the fit to the higher-resolution and more detailed ground-based
spectroscopy now available.  Even the Gauss-Hermite coefficients $h_3$
and $h_4$ -- which were not used in constructing the model -- were
adequately well fitted.  These models were then used to predict the
kinematics that should be observed in our Ca triplet HST spectra of
the red stars.  This is a stringent test because the new models were
used to predict observations taken at much higher resolution than
those used to construct the models.  Excellent fits were obtained.
This is a resounding success of the eccentric disk model.  The
structural and velocity asymmetries of the nucleus can be explained
almost perfectly if the eccentric disk is inclined with respect to the
plane of the outer disk of M{\ts}31.  Here, we publish the kinematic
data used by Peiris \& Tremaine (2003) in the above comparison
(\S\ts3), and we revisit particularly interesting features of the STIS
kinematics of P1 $+$ P2 in the Appendix.

      The main shortcoming of the Peiris \& Tremaine models is that they
do not include the self-gravity of the stars in the eccentric disk.~If 
the disk has a mass of 10\thinspace\% of the~BH,~then self-gravity is 
needed to keep the model aligned (Statler 1999).  The most detailed 
such models are by Salow \& Statler (2001, 2004).  They model all 
available observations but do not fit the data as well as the models 
by Peiris \& Tremaine (2003).  Other self-consistent models are based
on $N$-body simulations (Bacon et al.~2001, Jacobs \& Sellwood 2001); 
again, they reproduce only some of the observations.  Sambhus \& 
Sridhar (2002) use the Schwarzschild (1979) method to model the double 
nucleus.  The above models differ in many details.  For example, the 
Salow and Statler models precess
rapidly, with pattern speeds of $36 \pm 4$~km~s$^{-1}$~pc$^{-1}$; the 
models of Sambhus \& Sridhar precess at 16~km~s$^{-1}$~pc$^{-1}$, and
the simulations of Bacon et al.~(2001) precess at only 
3~km~s$^{-1}$~pc$^{-1}$.  Not surprisingly, the construction of 
dynamical models that include self-gravity is a challenge.  The 
conclusion that such models are long-lived is less secure than the 
result that they can instantaneously fit the photometry and kinematics 
of P1\ts$+${\ts}P2.  Tremaine (2001) gives a general discussion of 
slowly precessing eccentric disks.

      Because of these complications, the BH mass in M{\ts}31 has 
remained uncertain. Estimates of $M_\bullet$ by Dressler et al.~(1988), 
Kormendy (1988), Richstone et al.~(1990), Bacon et al.~(1994), Magorrian
et al.~(1998), KB, Bacon et al.~(2001), Peiris \& Tremaine (2003), and 
Salow \& Statler (2004) have ranged over a factor of about 3, 
$M_\bullet \simeq (3~\rm to~10) \times 10^7$~$M_\odot$.  These results 
are reviewed and error bars are tabulated on a uniform distance scale
($D = 0.76$ Mpc) in Kormendy (2004).  In this paper we show that an 
analysis of the UV-bright nucleus P3 allows us to estimate the black 
hole mass independent of P1~$+$~P2.

\vskip 100pt

\section{STIS Spectroscopy}

      The STIS CCD observations of M{\thinspace}31 were obtained on
1999 July 23 -- 24.  The slit was aligned at P.A. = $39\arcdeg$.
Other details of the STIS configuration are given in Table~1.  We
obtained a spectrum that includes the calcium triplet,
$\lambda\lambda$~8498, 8542, and 8662 \AA,~and one at
$\lambda\lambda$~2700 -- 5200~\AA~that includes several Balmer lines
and Ca II H and K ($\lambda\lambda$~3933 and 3968 \AA).  Both
wavelength regions were observed because we wanted separately to
analyze the double nucleus P1 $+$ P2 and the central blue cluster P3.
Figure 1 shows that the double nucleus P1 $+$ P2 contributes almost
all of the light at red wavelengths, while P3 dominates at 3000\ts\AA.
The color difference between P1 $+$ P2 and P3 is illustrated further
in the brightness cuts in Figure 3.  The red spectrum was obtained
using the $52 \times 0\farcs1$ slit, while the blue spectrum was taken
with the $52 \times 0\farcs2$ slit.  A wider slit was chosen for the
blue spectrum to ensure that P3 would fall inside the
slit.  Figure 1 shows the placement of the slit relative to the WFPC2
F555W and F300W images from Lauer et al.~(1998).  These slit positions
were determined by comparing the light profiles along the slit in our
STIS spectra with brightness cuts through the WFPC2 images.  We
measured the slit positions to an accuracy of $0\farcs005$ for the red
spectrum and $0\farcs01$ for the blue spectrum.  The total integration
time for the red spectrum was split into two exposures of
approximately 1200 s each per HST orbit.  M{\thinspace}31's nucleus
was shifted by 4.1 pixels along the slit between orbits.

      The total integration time for the blue spectrum was
split into three equal exposures within one HST orbit.  The nucleus
was shifted by 4.3 pixels between successive exposures.  Wavecals were
interspersed among the galaxy exposures to allow wavelength
calibration, including correction for thermal drifts.  For the red
spectrum, we obtained contemporaneous flat-field exposures through
the same slit while M{\thinspace}31 was occulted by the Earth.  These
provide proper calibration of internal fringing, which is significant
at $\lambda \geq 7500$~\AA \ (see Goudfrooij, Baum, \& Walsh 1997).

      The spectra were reduced as described in Bower et al.~(2001).
Unlike red spectra taken at $\lambda \geq 7500$~\AA~with the G750M
grating, blue spectra taken with the G430L grating are not affected by
fringing.  Consequently, we flat-fielded the G430L data using the
library flat image from the STScI archive.  The final reduced spectra
have maximum signal-to-noise values of $S/N = 25$ \AA$^{-1}$ \ (G750M)
and $50$ \AA$^{-1}$ (G430L).

      A stellar template spectrum is needed to measure the stellar
kinematics implied by the galaxy spectra.  For the red spectrum of
M{\thinspace}31, our template is the STIS spectrum of HR~7615 from
Bower et al.~(2001).  They document the observational setup and data
reduction for this spectrum. For the blue spectrum we used template A
stars from Le Borgne et al.~(2003), white dwarf stars observed in the
Sloan Digital Sky Survey (Kleinman et al. 2004) or modeled by Finley,
Koester and Basri (1997) and Koester et al. (2001), and spectral
syntheses of various stellar population models by Bruzual \& Charlot
(2003).  These sources were supplemented for checking purposes by using
standard stars from Pickles (1998).  Spectral resolution is not an
issue for standard stars, because the intrinsic width of the
absorption lines in A-type stars is much larger than the instrumental
width of STIS with the G430L grating, and because the spectrum of P3
proves to have exceedingly broad lines.

\centerline{\null} \vfill


\begin{figure*}[t]
\centerline{\psfig{file=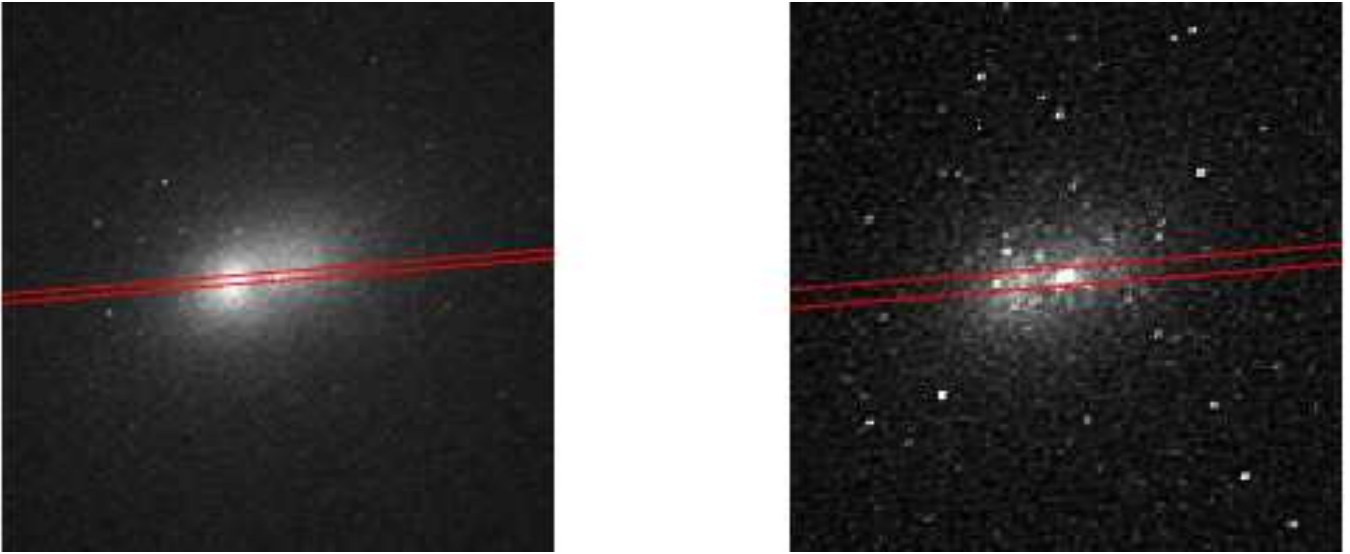,width=18cm,angle=0}}
\figcaption[bender.fig1.ps]{STIS slit positions superimposed on the WFPC2 
images from Lauer et al.~(1998). The left panel shows the 0\farcs1 slit 
position for the Ca II spectrum on the F555W image, and the right panel shows
the 0\farcs2 slit position for the blue spectrum on the F300W image. The 
images cover the central 6.4 arcsec by 6.4 arcsec. North is 55.7 degrees
counter-clockwise from up.  \label{}}
\end{figure*}



\begin{deluxetable}{lll}
\tablenum{1}
\tablewidth{460pt}
\tablecaption{STIS Instrument Configurations }
\tablehead{
\colhead{Parameter~~~~~~~~~~~~~~~~~~~~~~~~~~~~~~~~~~~~~~~~~~~~~~~~} & 
\colhead{Red Spectrum~~~~~~} & 
\colhead{Blue Spectrum~~~~~~}}
\startdata
Detector gain (e$^{-}$ per ADU) & 1.0 & 1.0 \nl
Grating & G750M & G430L \nl
Wavelength range & 8272 \AA \ $-$ 8845 \AA & 2900 \AA \ $-$ 5700 \AA \nl
Reciprocal dispersion (\AA \ pixel$^{-1}$) & 0.56 & 2.73 \nl
Slit width (arcsec) & 0.1 & 0.2 \nl
Comparison line FWHM (pixel) & 3.1 & 3.5 \nl
R $= \lambda / \Delta \lambda $ & 4930 & 450 \nl
Instrumental dispersion $\sigma_{\rm instr}$ (km s$^{-1}$) & 56 & 284 \nl
Scale along slit (arcsec \ pixel$^{-1}$) & 0.051 & 0.051 \nl
Slit length (arcmin) & 0.8 & 0.8 \nl
Integration time (sec) & 20790 & 2040 \nl
\enddata
\end{deluxetable}

\vspace*{.5cm}


\section{Kinematics of the Double Nucleus P1 $+$ P2}

      The calcium triplet spectroscopy sees only the red giant stars
that make up the double nucleus, P1 $+$ P2.  It is blind to P3,
which contributes essentially no light at $\lambda \simeq 8500$~\AA.
The kinematic properties of the red stars are illustrated in Figure 2.

      In Figure 2, the spectrum of the bulge has been subtracted
following procedures discussed in KB.  Bulge subtraction is analogous
to sky subtraction in the sense that it removes the effects of a
contaminating spectrum that is not of present interest.  As shown in
KB, the bulge of M{\ts}31 dominates the light distribution only at
radii $r$ \gapprox \ts 2$^{\prime\prime}$. At $r < 1^{\prime\prime}$,
it contributes about 20\ts\% of the light.  So over the radii of
interest in Figure 2, bulge stars are a minor foreground and
background contaminant; they do not significantly participate in the
dynamics of the double nucleus.  It is routine to estimate the small
contribution of bulge stars to the STIS red spectrum and to subtract
it.  Figure 2 is therefore a pure measure of the kinematics of the
stars that make up the double nucleus.

\vskip 10pt

      Figure 2 also shows the bulge-subtracted nuclear kinematics
measured with the Canada-France-Hawaii Telescope (CFHT) (KB). Taking
into account both the PSF and the slit, the effective Gaussian
dispersion radius of the effective PSF was $\sigma_* = 0\sd297$
(Kormendy 2004).  The corresponding resolution of the STIS red
spectroscopy is $\sigma_* = 0\sd052$.

      Confirming results of KB, the dispersion profile of the red stars 
reaches a sharp peak slightly on the anti-P1 side of P3.  The peak
dispersion is higher at STIS resolution ($\sigma = 373 \pm 50$ km s$^{-1}$)
than at CFHT resolution ($\sigma = 287 \pm 9$ km s$^{-1}$).  The rotation
curve is also asymmetric; the maximum rotation velocity is larger on the
anti-P1 side than it is in P1.  Again, the asymmetry is larger and the
radius of maximum rotation is smaller at STIS resolution than at CFHT
resolution.  These observations are consistent with and provide further
evidence for Tremaine's (1995) model for the double nucleus as an eccentric 
disk of stars orbiting the central BH. The Appendix provides more detailed
discussion.

 

\psfig{file=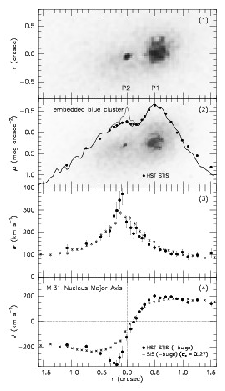,width=8.3cm,angle=0} \figcaption[]{Panel 1
shows the double nucleus of M{\thinspace}31 rotated $\sim 185^\circ$
clockwise with respect to Figure 1.  It is a
$I$\ts$+$\ts$V$\ts$+$\ts3000 \AA\ composite from KB.  P1 is brighter
than P2 in red light.  Embedded in P2 is P3, i.{\ts}e., a tiny cluster
of blue stars that is invisible in $I$ but brighter than P1 in the
ultraviolet.  The background image in Panel 2 is a similar
$V$\thinspace$+$\thinspace\thinspace3000 \AA\ composite that better
shows the small radius of P3. Panel 2 includes an $I$-band brightness
cut along the P1\ts--\ts{P2} axis (lower~curve) and a $V$-band cut
through the blue cluster P3 (upper curve). The points are the
brightness profile in the STIS spectrum; they are used to register the
kinematics with the photometry in radius.  Along the P1\ts--\ts{P2}
axis, radius $r = 0$ is chosen to be the center of P3 (note that in
KB we centered the radius scale at $0\farcs068$, not P3).  Panels 3
and 4 show velocity dispersions and rotation velocities along the
\hbox{P1\ts--\ts{P2}} axis after subtraction of the bulge.  The
ground-based points (crosses) are from the Subarcsecond Imaging
Spectrograph (``SIS'') and the CFHT (KB).  The STIS data (filled 
circles) are Fourier quotient reductions.  Bacon et~al.~(2001) made 
an independent reduction of our red STIS spectrum; it is consistent 
with ours.  
\label{}}

\centerline{\null}


\section{The Integrated Spectrum of P3}

\subsection{P3 is Made of A-Type Stars}

     P3, the compact blue cluster, is illustrated in the two panels of
images in Figure 2.  It is embedded in P2 but is not concentric with
it; the photocenter of P2 is $\sim$\ts0\farcs03 on the anti-P1 side of
the blue cluster. The center of the bulge is slightly off in the
opposite direction, i.{\ts}e., toward P1 (see KB and discussion
below).  Note that we choose $r = 0$ to be the center of P3, whereas
KB chose $r = 0$ to be the center of the bulge.


\psfig{file=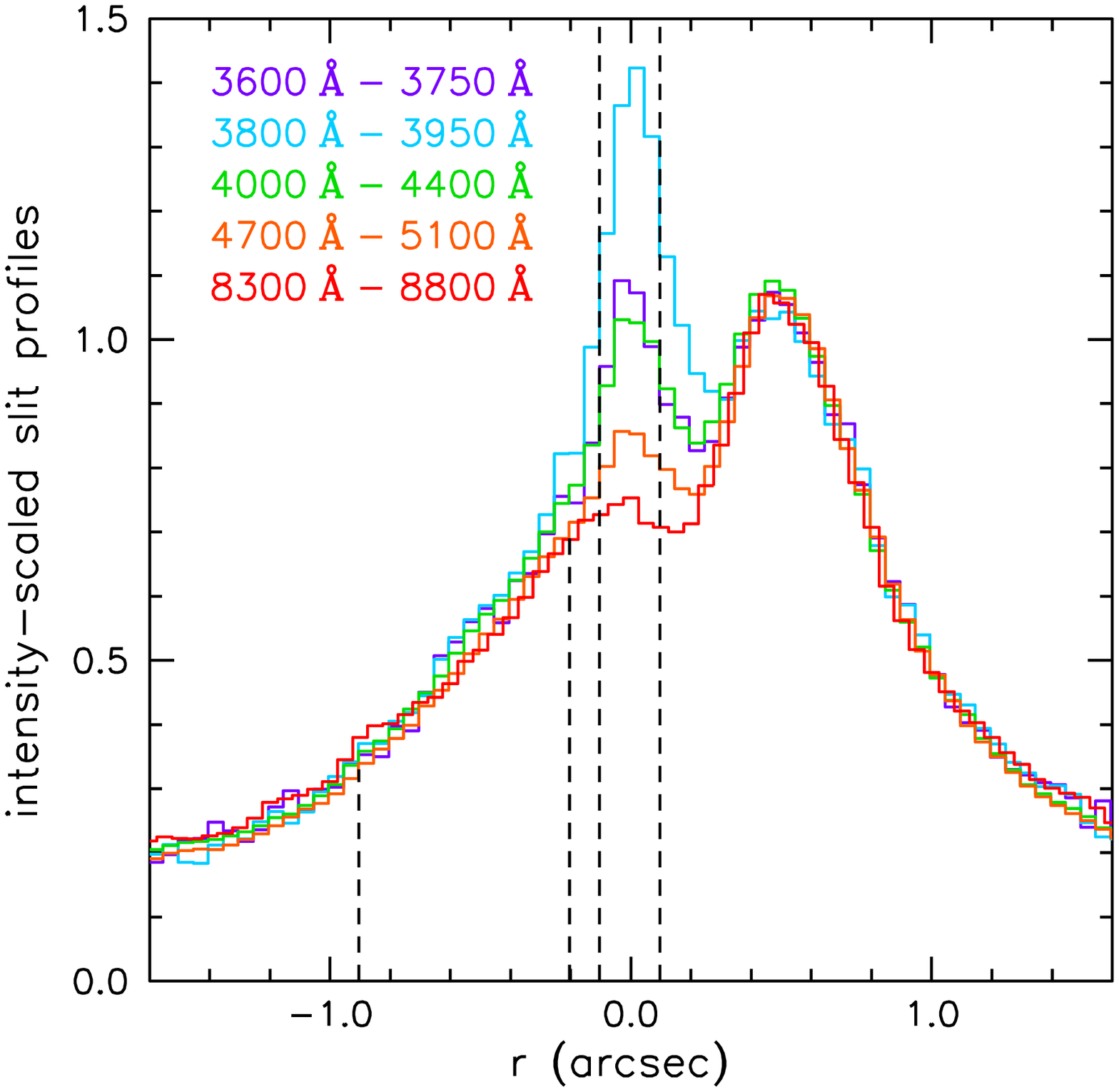,width=9cm,angle=0}
\figcaption[]
{Linear intensity cuts through the blue and red spectra of
P1\ts$+$\ts{P2}\ts$+$\ts{P3}.  Each cut is an average over the wavelength range
given in the key. The contrast between the blue cluster P3 and the
underlying red nucleus P2 is largest at 4000\ts\AA.  It is smaller at redder
wavelengths because the stars in P3 are blue.  It is smaller at bluer 
wavelengths because the spectrum of P3 has a strong Balmer break (Figure 4). 
The two leftmost vertical dashed lines indicate the region in which the
background spectrum was derived. The two rightmost vertical dashed lines
indicate the radius range over which we averaged the background-subtracted P3
spectrum shown in Figures 4, 5 and 6.  
\label{}}
\vskip -0.04cm


      We obtained our STIS spectrum at $\lambda\lambda \simeq 3500$ to
5000~\AA\ in part to study this issue.  Over the above wavelength
range, P3 provides a strong signal, much stronger than that indicated
by the $V$-band brightness cut in Figure~2.  Figure 3 shows brightness
cuts through the red and blue STIS spectra in various wavelength
ranges. The blue cluster is essentially invisible at
8300\ts--{\ts}8800 \AA\ in the red spectrum.  We assume that this
spectrum provides the surface brightness profile of the underlying
double nucleus.  With respect to this profile, P3 is, in general, more
prominent at bluer wavelengths.\footnote{P3 looks fainter at
4700\ts--{\ts}5100 \AA\ in Figure 3 than at 5500\ts\AA\ ($V$ band) in
Figure 2.  The reason is that Figure 2 shows a brightness cut through
the deconvolved $V$-band image from Lauer et al.~(1998); this has
higher spatial resolution than an undeconvolved STIS spectrum.  Also,
the $V$-band cut is 0\farcs046 wide, while the spectrum was obtained
through a 0\farcs2 wide slit.}  The contrast over P1 $+$ P2 is highest
at 3800\ts--\ts3950 \AA.  Then P3 gets less prominent at
3600\ts--\ts3750 \AA; the reason turns out to be that the spectrum has
a strong Balmer break (Figure 4).  The important conclusion from
Figure 3 is that the spectrum of P3 is almost as bright
as the underlying spectrum of P1 $+$ P2 at just the wavelengths where
hydrogen Balmer lines are strongest.

      It is therefore possible to extract a clean spectrum of P3 despite the
short integration time and modest signal-to-noise ratio.  We averaged the
spectrum of P3 over the 0\farcs2 = four spectral rows in which it is brightest
(right pair of dashed lines in Figure 3).  We approximated the spectrum of the
underlying P2 stars by averaging 14 rows of the spectrum on the anti-P1 side 
of P3 (left pair of dashed lines in Figure 3.)  The 8300\ts--\ts8800 \AA\ 
brightness cut was used to scale this average P2 spectrum to the P2 brightness
underlying P3.  The result was subtracted from the four-row average spectrum
of P2 $+$ P3.  The resulting spectrum of P3 is shown in black in Figures
4\ts--\ts6.


\begin{figure*}[t]
\centerline{\psfig{file=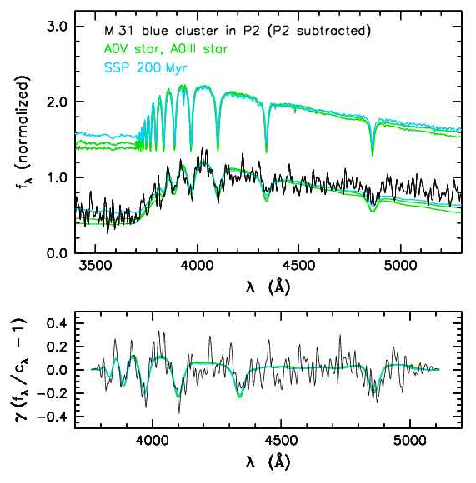,width=14cm,angle=0}}
\figcaption[]
{Spectrum (black) of the central 0\farcs2 of the blue cluster P3.  The superposed
spectrum of the stars in the bulge and nucleus has been subtracted.  Flux is in
arbitrary linear units.  In the lower panel, the spectrum has been divided by a
polynomial $c_\lambda$ fitted to the continuum; it has been normalized to zero
intensity, and multiplied by the mean ratio $\gamma$ of the line strength in 
the standard stars to that in P3.  The colored lines show the
spectra of an A0 dwarf star, an A0 giant star, and a Bruzual \& Charlot (2003)
starburst of age 200 Myr before (top) and after (overplotted on the data)
broadening to the line-of sight velocity distribution that best fits the 
cluster spectrum.  The fit was carried out with the Fourier correlation 
quotient program (Bender 1990).   
\label{}}
\end{figure*}
\vskip 0.6cm



\begin{figure*}[t]
\centerline{\psfig{file=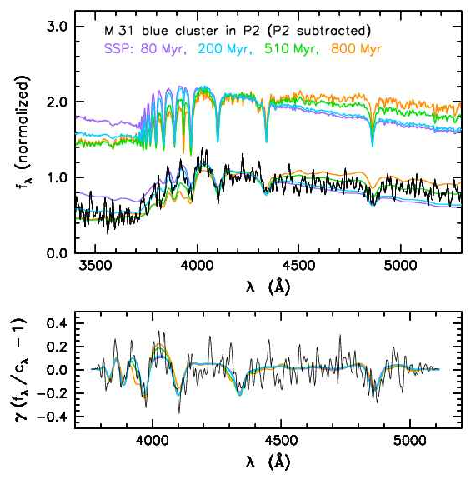,width=14.4cm,angle=0}}
\figcaption[]{This figure is analogous to Figure 4 except that the
spectrum of the blue cluster P3 is compared with Bruzual \& Charlot
(2003) starbursts of various ages given in the key.  The fit to the
red continuum is best for an age of $\sim$ 510 Myr, but then the
strengths of the Balmer lines H$n$ for $n \geq 5$ are wrong relative
to the strengths of the redder lines.  This problem gets worse for
older starburst ages.  Starbursts younger than 200 Myr are too blue;
their Balmer breaks are too small to fit the observed spectrum.
\label{}}
\end{figure*}
\vskip 0.3cm


      The stellar population of P3 is dramatically different from that
of P1 and P2.  The spectrum in Figure 4 is dominated by Balmer
absorption lines.  At least five Balmer lines are visible, starting
with H$\beta$ at $\lambda_{\rm obs} \simeq 4856$\ts\AA.  Also
prominent is a strong Balmer break.  In fact, the spectrum is very
well matched by velocity-broadened spectra of A giant and dwarf stars.
This confirms that the nucleus is made mostly of A-type stars as Lauer et
al. (1998) and Brown et al. (1998) suggested.

\subsection{The Remarkably High Velocity Dispersion of P3:\\
            The Supermassive Black Hole Is In The Blue Cluster}

      The blue cluster has a remarkably high velocity dispersion.
Using an A0 dwarf star from Le Borgne et al.~(2003) as a template, the
Fourier correlation quotient program (Bender 1990) gives a velocity
dispersion of $\sigma = 962 \pm 105$ km s$^{-1}$.  An A0 giant star
gives $\sigma = 984 \pm 107$ km s$^{-1}$.  A-type stars have
intrinsically broad lines, but $\sigma$ is so large that the difference 
between using giants and dwarfs is insignificant.  The above fits are
illustrated in Figure 4.  The match to the lines and to the Balmer break 
is excellent.  The results are robust; plausible changes in the intensity
scaling of the P2 spectrum that was subtracted produce no significant change
in $\sigma$.  

      The best-fitting 200-Myr-old stellar population model (Figure 5, \S\ts4.3)
gives a dispersion of $\sigma = 984 \pm 106$ km s$^{-1}$.
We adopt the average of the dispersion values given by the A dwarf star, the
A giant star, and the 200-Myr-old stellar population model; this gives 
$\sigma = 977 \pm 106$ km s$^{-1}$ as our measure of the velocity dispersion 
of P3 integrated over the central 0\farcs02.  

      Despite its tiny size (half-power radius $\simeq 0\farcs06 \simeq 0.2$ 
pc; Lauer et al.~1998), P3 has the highest integrated velocity dispersion
measured to date in any galaxy.  The velocity dispersion of P3 is even larger
than the line-of-sight velocity dispersion of the Sgr A* cluster in our Galaxy
($\sigma = 498 \pm 52$ km s$^{-1}$ to $840 \pm 104$ km s$^{-1}$, depending on
the sample of stars chosen, Sch\"odel et al.~2003)\footnote{Of course, the 
pericenter velocities of the innermost individual stars in our Galaxy are in
some cases much larger.  The current record is held by S0-16, which was moving
at $12000 \pm 2000$ km s$^{-1}$ when it passed within 45 AU = 0.0002 pc = 600
Schwarzschild radii of the Galaxy's BH (Ghez et al.~2004).}. 
The high velocity dispersion of P3 is especially remarkable in 
view of the observation (Figure 2) that the velocity dispersion of the red 
stars {\it along the same line of sight} is only $\sim 250$ km s$^{-1}$.  
The maximum velocity dispersion of P2, $373 \pm 48$ km s$^{-1}$ at $\Delta r 
\simeq 0\farcs06$ on the anti-P1 side of the blue cluster, is much smaller than
that of P3.  Even the remarkably high velocity dispersion, $\sigma = 440 \pm
70$ km s$^{-1}$ measured in P2 by Statler et al.~(1999) is much smaller than
the velocity dispersion of P3.  This confirms the conclusion 
of KB that the M{\ts}31 supermassive black hole is in the blue cluster.

\subsection{Fit of a Starburst Spectrum to P3}

      The overall continuum slope of P3 is best fitted not by a single
A-type star but rather by the spectrum (Bruzual \& Charlot 2003) of a
single starburst population (SSP in Figures 4 and 5) of age $\sim 200
\pm 50$ Myr and solar metallicity.  The blue continuum fit is
essentially perfect; the red continuum fit is improved slightly over
the single star fits.  Starburst spectra with a range of ages are
shown in Figure~5.  Using older starbursts allows us to fix the fit to
the \hbox{5000-\AA} continuum, but only at an unacceptable price: the
bluest Balmer lines are no longer well fitted.  Complicating the model
further would be overinterpretation; the error in the red continuum
fit could be due to imperfect P2 subtraction or to small
amounts~of~dust. But it is clear that we cannot exclude some
admixture of older stars.  Reasonable changes in metallicity also do
not affect the fit: metallicity changes are largely degenerate with
age changes.


\psfig{file=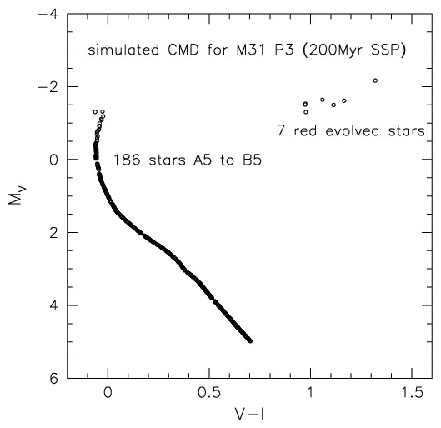,width=7.0cm,angle=0} 
\figcaption[]{Sample colour magnitude diagram of a 200 Myr old
single burst population with solar metallicity and a total luminosity of
$M_V = -5.7$. The spectrum is
dominated by stars of $\sim$~10000~K temperature. The diagram has been
generated using the synthetic color-magnitude diagram algorithm of the
Instituto de Astrofisica de Canarias (Aparicio \& Gallart 2004).
\label{}}
\vskip 0.3cm


\vskip 3pt

      How many stars make up P3?  For an absolute visual magnitude of
$M_V \approx -5.7$ (Lauer et al.~1998), we estimate the answer in
Figure 6, using IAC-STAR, the synthetic color-magnitude diagram (CMD)
algorithm of the Instituto de Astrofisica de Canarias (Aparicio \&
Gallart 2004). A 200-Myr-old, single-burst population of solar
abundance implies that about 200 stars between spectral types A5 and
B5 dominate the spectrum.  The large number of stars at the same
temperature of $\sim$\ts10000 K explains why the spectrum of P3 is so
similar to that of a single A0 star.  Figure 6 also shows why P3 has a
fairly smooth appearance, although surface brightness fluctuations are
visible in Figure 8.  Only a few red evolved stars are present, and
they do not contribute significantly to the light of P3. Future
observations with resolutions of about 0.01 arcsec should resolve the
brightest stars close to the BH.

      For a Salpeter (1955) initial mass function with a lower mass cut-off 
at 0.1 $M_\odot$, the total number of stars on the main sequence at present
is $\sim$ 15000, their total mass is about 4200 solar masses. If the
burst originally produced stars up to 100 $M_\odot$, then the initial
total mass of P3 was $\sim$ 5200 $M_\odot$. Given the inefficiency of 
star formation, the total gas mass required to form P3 probably was of the
order $10^6$ $M_\odot$.

      Forming stars so close to a black hole is not trivial.  It may
be possible if $\sim 3 \times 10^6{\ts}M_\odot$ of gas could be
concentrated into a thin disk of radius 0.3 pc and velocity dispersion
10~km~s$^{-1}$.  Then Toomre's (1964) stability parameter Q $\simeq$
1.  It is not easy to see how such an extreme configuration could be
set up, especially without forming stars already at larger radii.
Well before the black hole makes star formation difficult, the surface
density of the dissipating and shrinking gas disk would get high
enough so that the Schmidt (1959) law observed in nuclear starbursting
disks (Kormendy \& Kennicutt 2004, Figure 21) would imply a very high
star formation rate.  This star formation would have to be quenched
until the gas disk got small enough to form P3.  And then the star
formation would have to be very inefficient to put only $\sim$ 5200
$M_\odot$ of the $\sim 3 \times 10^6{\ts}M_\odot$ of gas into stars.
Similar considerations make it difficult to understand young stars
near the Galactic center black hole (e.{\ts}g., Morris 1993; Genzel et
al. 2003, Ghez et al.~2003, 2004).  Nevertheless, young stars -- or at
least:~high-luminosity, hot stars -- are present.  Complicated
processes of star formation (e.{\ts}g., Sanders 1998) may not
realistically be evaluated by a simple argument based on the Toomre
$Q$ instability parameter.  So, if a dense enough and cold enough gas
disk can be formed, star formation may be possible, even close to a
supermassive black hole.

\centerline{\null}

\subsection{Could the Hot Stars in P3 Result From Stellar Collisions?}

      The alternative to a starburst could be that the hot stars 
of P3 are formed via collisions between lower mass stars in P3 or even in
P1\ts$+${\ts}P2.  Yu (2003) argues that the collision timescales are too
long to be of interest.  It would be interesting to revisit this issue 
given the conclusion of \S\thinspace6.1 that P3 is a cold stellar disk.
In any case, it is worth noting that the conversion of (say) a high-mass,
0.5  $M_\odot$ main sequence star in P2 into an A star requires
merging $\sim 6$ stars without mass loss.  It is not easy to see how
the A stars in P3 could originate by collisions.

      Thus the situation in P3 is similar to that in our Galaxy.  
No explanation of the hot stars looks especially plausible.

\centerline{\null}
 
\subsection{P3 is Not Made of White Dwarf Stars}

      Finally, we need a sanity check to make sure that we are not 
completely misinterpreting the observations.  Dynamically, we detect a 
\hbox{$10^8$-$M_\odot$} central dark object.  This is associated with 
a tiny and faint nucleus comprised of hot stars that have extraordinarily broad 
absorption lines.  White dwarf stars have extraordinarily broad absorption
lines.  If they are not too old, they can easily have an A-type spectrum, 
and if they are not too young, they can easily contribute mass without
contributing much light.  It is natural to wonder -- could P3 be a cluster 
of white dwarfs?  Could they simultaneously explain the broad-lined, 
A-type spectrum and the central dark mass?  This possibility is not
excluded by stellar collision or cluster evaporation timescales
(Maoz 1995, 1998). 

      Figures 7 and 8 show that P3 cannot be made of white dwarfs.
Figure 7 compares the spectrum of P3 with that of a typical DA white 
dwarf observed in the Sloan Digital Sky Survey (SDSS).  The star was 
chosen to have Balmer line strengths comparable to those in P3.  It is
approximately the best match to P3 that can be achieved with white 
dwarf spectra.  Its lines are narrower than those of P3, so we can fit
the observed line widths (bottom panel of Figure 7) with 
$\sigma = 885 \pm 126$ km s$^{-1}$.  That is, this relatively
narrow-lined white dwarf gives a dispersion similar to those implied by
main sequence and giant A stars.  The fit to the line widths is less 
good than the fit provided by A0{\ts}V stars, but it is not inconsistent
with our low $S/N$ spectrum of P3.  If we had only the spectrum of this 
white dwarf as observed over the relatively narrow wavelength region 
redward of the Balmer break, we could not exclude white dwarf stars.

      However, the continua of white dwarf stars do not fit the large
Balmer break in P3.  SDSS J094624.30$+$581445.4 (Figure 7)
does not show this -- it and most other white dwarfs have not been 
observed at blue enough wavelengths to reach the Balmer break.  
Therefore we resort to model spectra kindly provided by Detlev Koester
(Finley, Koester and Basri 1997; Koester et al.~2001).  Figure 7 shows
a model spectrum that has line profiles similar to those in the 
observed white dwarf.  The price of having narrow enough lines to fit
the absorption lines in P3 is that there is essentially no Balmer break. 
Such a star cannot fit the continuum of P3.  This result is very robust;
it is not affected by uncertainities in the subtraction of the spectra 
of P1 $+$ P2.

\vskip 0.3cm

\begin{figure*}[t]
\centerline{\psfig{file=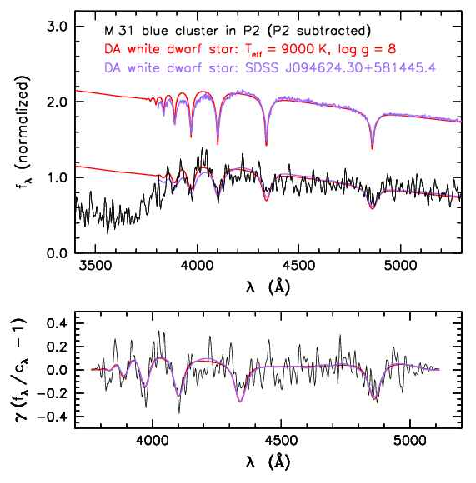,width=14cm,angle=0}}
\figcaption[m31_blueclus_spec_whitedwarf.cps]{Spectrum of P3 fitted with
approximately matched observed and model white dwarf spectra.  The observed
spectrum is from the SDSS (Kleinman et al.~2004; see http://www.sdss.org).
The model spectra used in Figures 7 and 8 are from Finley, Koester, \&
Basri (1997) and from Koester et al.~(2001).  The fits of white dwarf
spectra to P3 are significantly worse than the ones in Figures 4 and 5.  
The absorption lines of the white dwarfs are intrinsically too strong, and 
the white dwarfs fail completely to fit the large Balmer continuum break 
in the P3 spectrum.  However, the implied velocity dispersion, 
$\sigma = 885 \pm 126$ km s$^{-1}$, is consistent with our adopted value, 
$\sigma = 977 \pm 106$ km s$^{-1}$.
\label{fig7}}
\end{figure*}
\vskip 0.3cm

   Choosing different white dwarf parameters does not solve this problem.
No combination of temperature and gravity allows a simultaneous match to 
the Balmer line strengths and the Balmer break.  Figure 8 shows fits of
model white dwarf spectra with temperatures $T = 7000$ K, 8000 K, 10000 K,
and 12000 K, respectively.  For each temperature, we try surface gravities
of $10^7$, $10^8$, and 10$^9$ cm s$^{-2}$.  

      Temperature $T \simeq 7000$ K is too cold.  The stellar lines are
too weak.  Not surprisingly, these stars have no Balmer break at all.  
Despite the bad continuum fit, the narrow lines in the white dwarf
templates give dispersions, $\sigma = 945 \pm 103$ km s$^{-1}$, 
$\sigma = 987 \pm 107$ km s$^{-1}$, and $\sigma = 1063 \pm 115$ km s$^{-1}$,
that are consistent with our adopted result.

      At $T = 8000$ K, the fit to the lines is better, although not as 
good as for A0{\ts} dwarf or giant stars.  The dispersion remains high 
($\sigma = 930 \pm 101$ km s$^{-1}$, $\sigma = 929 \pm 101$ km s$^{-1}$, 
and $\sigma = 952 \pm 103$ km s$^{-1}$).  Again, the Balmer break in the 
white dwarfs is too weak.

      At $T = 10000$ K, the stellar lines are much broader.  
The fit to P3 is acceptable after scaling the line strengths.  
For $\log {g} = 7$, 8, and 9, $\sigma = 784 \pm 120$ km s$^{-1}$, 
$769 \pm 134$ km s$^{-1}$, and $821 \pm 150$ km s$^{-1}$, respectively. 
Note that without line-strength scaling, the broadened white dwarf spectrum 
does not fit the galaxy.  And, even though the lines are now strong enough
when $\log {g}$ is large to produce a Balmer break, it is still too small 
to fit the spectrum of P3.  The green line emphasizes how much an 
A0{\ts}V star fits the spectrum of P3 better than does any white dwarf.

      Increasing the temperature further is counterproductive.  At 
$T = 12000$ K, the lines are too strong and too broad to fit P3, although 
we still obtain high dispersions ($\sigma = 705 \pm 144$ km s$^{-1}$,
$\sigma = 676 \pm 166$ km s$^{-1}$, and $\sigma = 680 \pm 190$ km s$^{-1}$).
Even high temperatures do not produce strong enough Balmer breaks.

      We conclude that no spectral synthesis of white dwarf stars of
different temperatures or gravities would fit P3.  The ones that fail
least badly -- those that fit the lines but not the Balmer break --
imply velocity dispersions that are consistent with values derived from
A0 dwarf or giant stars.

      For a compact cluster of white dwarfs to be a viable alternative
to a supermassive black hole, it must be dark.  That is, it must be old. 
We explore this option further in Section \S\ts8.


\begin{figure*}[t]
\centerline{\hbox{\psfig{file=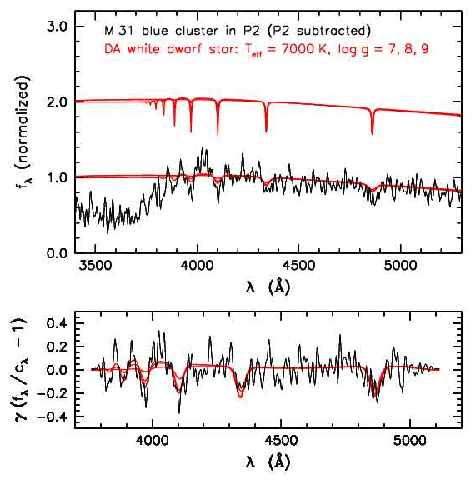,width=8.8cm,angle=0}
\psfig{file=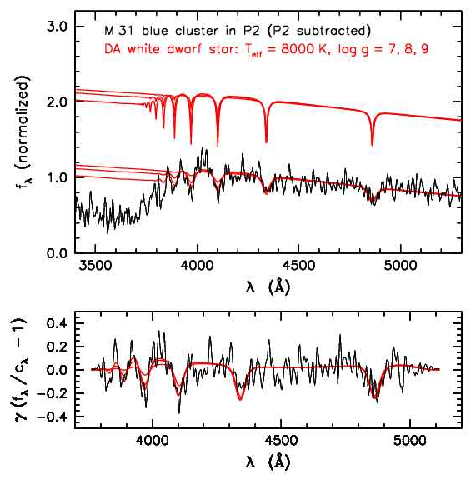,width=8.8cm,angle=0}}}
\centerline{\hbox{\psfig{file=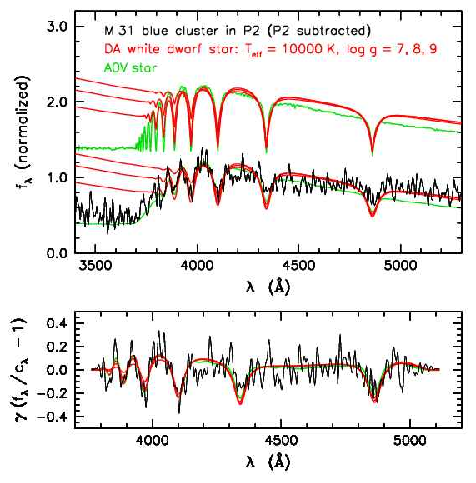,width=8.8cm,angle=0}
\psfig{file=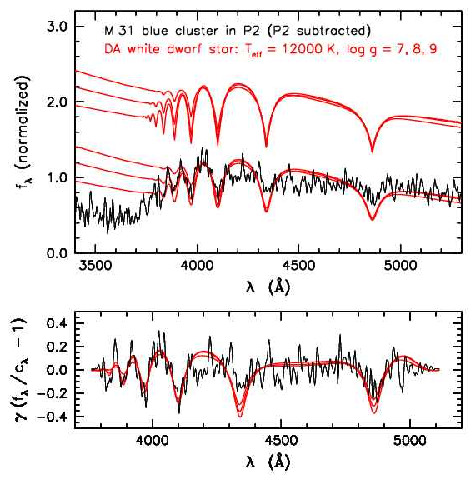,width=8.8cm,angle=0}}}
\figcaption[m31_figwd08.cps]{
Fits to the spectrum of P3 (black) of model white dwarf spectra (red) 
with temperature $T_{\rm eff} = 7000$ K, 8000 K, 10000 K, and 12000 K
(see the key).  At each temperature, surface gravities of $g = 10^7$, 
$10^8$, and $10^9$ cm s$^{-2}$ are used.  The green line shows the fit 
of an A0 dwarf star.  Compare Figure 6.
\label{xxx}}
\end{figure*}
\vskip 0.4cm

\section{Light Distribution of P3}

\vskip 2pt

      For a dynamical analysis of P3 (\S\ts6), we need its light
distribution with P1{\ts}$+${\ts}P2 subtracted.  To derive this, we
scaled the HST F555W image to the HST F300W image such that P1
disappeared after subtraction. The resulting image of P3 is shown in
Figure 9.  We then fitted P3 with S\'ersic (1968) models,
$$ I(r) = I_0 \exp[-(r/r_0)^{1/n}]~, $$ convolved with the HST point
spread function as in Lauer et al.~(1998).  The PSF was constructed
from two exposures of the

\centerline{\null}

\centerline{\null}

\centerline{\null}

\centerline{\null}

\noindent standard star GRW+70D5824 (u2tx010at,
u2tx020at). The free parameters in the fit were central surface
brightness $SB_0$, scale length $r_0 = \sqrt{a_0 b_0}$ ($a$, $b$ =
semimajor, semiminor axis), S\'ersic $n$, position angle $P.A.$,
ellipticity $1 - b/a$, and center coordinates.  Individual faint
point-like sources in the outskirts of P3 were masked before fitting.
The best fit over the radius range $r < 0\farcs3$ was obtained for
S\'ersic index $n = 1$, major-axis scale length $a_0 = 0\farcs1 \pm
0\farcs01$, PSF-convolved ellipticity $1-b/a = 0.33 \pm 0.03$, and
position angle $P.A. = 63^\circ \pm 2^\circ$ (this is 119$^\circ$
counterclockwise from vertical in Figure 9).


\begin{figure*}[t]
\centerline{\hbox{
\psfig{file=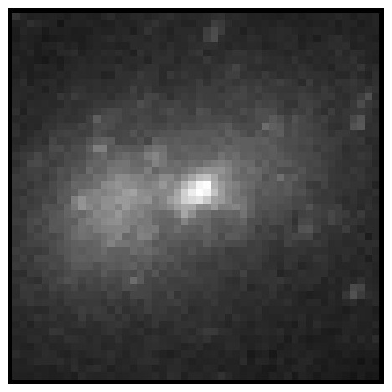,width=4.5cm,angle=0}
\psfig{file=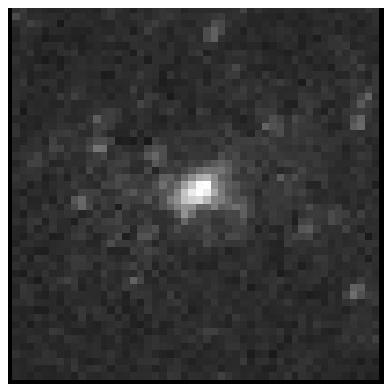,width=4.5cm,angle=0}
\psfig{file=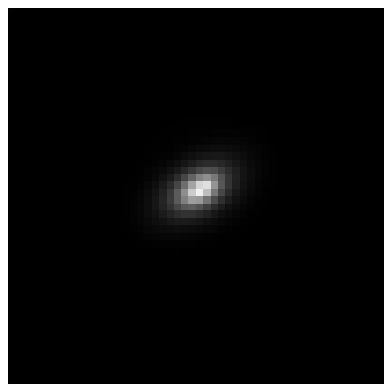,width=4.5cm,angle=0}
\psfig{file=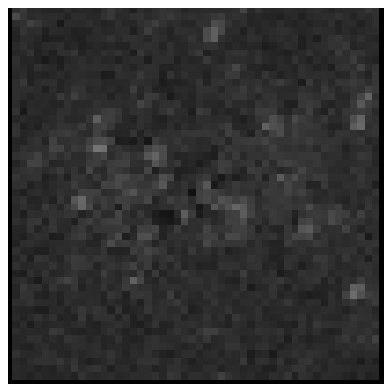,width=4.5cm,angle=0}}}
\figcaption[p3_image_f30vcndes.ps]{Left to right: (i) F300W image of the 
blue nucleus, P3, superposed on nuclei P1 and P2; (ii) F300W image of P3 
after subtraction of the F555W image intensity-scaled to the F300W image
in (i); (iii) PSF-convolved inclined disk model for P3; and (iv) difference 
between images (ii) and (iii) showing the quality of the model and the
residual surface brightness fluctuations.  All images are 2\farcs5 by 
2\farcs5. North is 55.7 degrees counter-clockwise from up, as in Figure 1.
\label{}}
\end{figure*}


\vskip 0.2cm
\centerline{\psfig{file=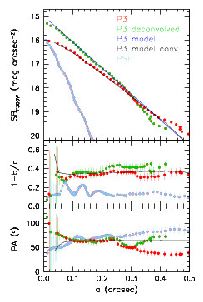,width=7.5cm,angle=0}}
\figcaption[p3_iso_nf.ps]{Observed radial profiles (red) of P3 surface
brightness SB, ellipticity $1-b/a$, and position angle PA versus
semi-major axis $a$.  Lucy-deconvolved profiles are shown in green.
The HST PSF is shown in light blue (with arbtitrary zeropoint).  The
inclined disk model before and after convolution with the HST PSF is
represented by blue and black lines, respectively.  The observed
profiles are over-sampled -- neighboring points are not independent.
\label{fig9}}

\centerline{\null}\vskip -1.0cm\centerline{\null}



\begin{center}
\renewcommand{\arraystretch}{1.05}

\textsc{Table 2}

\vskip 0.1cm 

\textsc{Parameters of the thin disk model of P3}

\vskip 0.2cm 

\begin{tabular}{ll}
\tableline
\tableline
Parameter\phantom{$^)_)$} & Value \\ 
\tableline 
$m_{\rm F300W}$      ~~~~~~~~~~~~~ & $18.6 \pm 0.1 $      \\
$M_{\rm F300W}$      ~~~~~~~~~~~~~ & $-5.8 \pm 0.1 $      \\
S\'ersic $n$                       & 1                      \\
exp. scale length $a_0$            & $0\farcs1 \pm 0\farcs01$ \\
$SB_{0,{\rm F300W}}$ (face-on)     & $15.6 \pm 0.1$ mag arcsec$^{-2}$ \\ 
inclination                        & $55^\circ \pm 2^\circ$ \\
position angle                     & $63^\circ \pm 2^\circ$ \\
\tableline
\end{tabular}
\end{center}




\vskip 0.3cm \centerline{\hbox{
\psfig{file=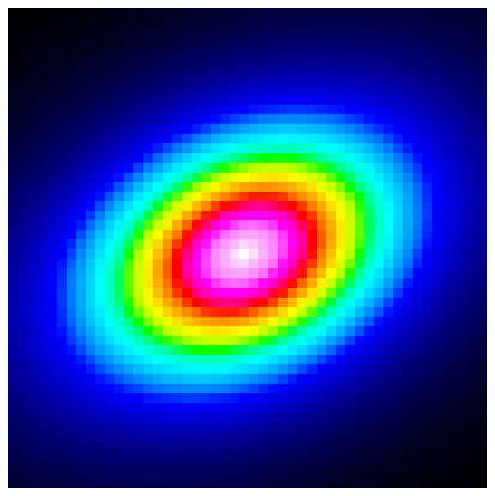,width=2.8cm,angle=0}
\psfig{file=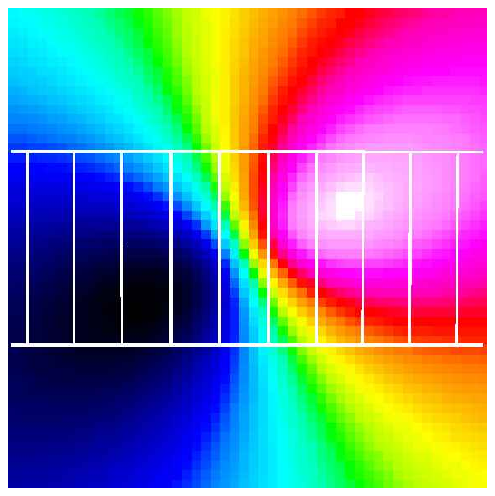,width=2.8cm,angle=0}
\psfig{file=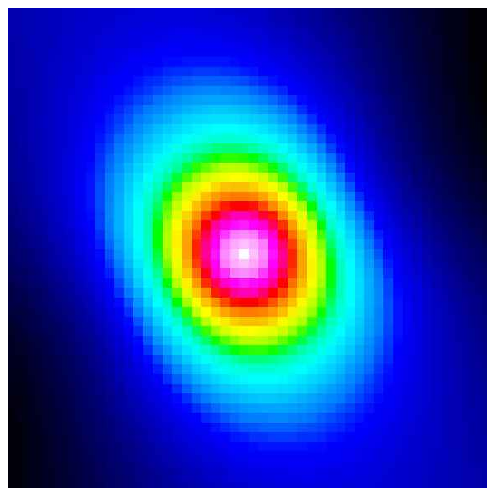,width=2.8cm,angle=0}}}
\centerline{\psfig{file=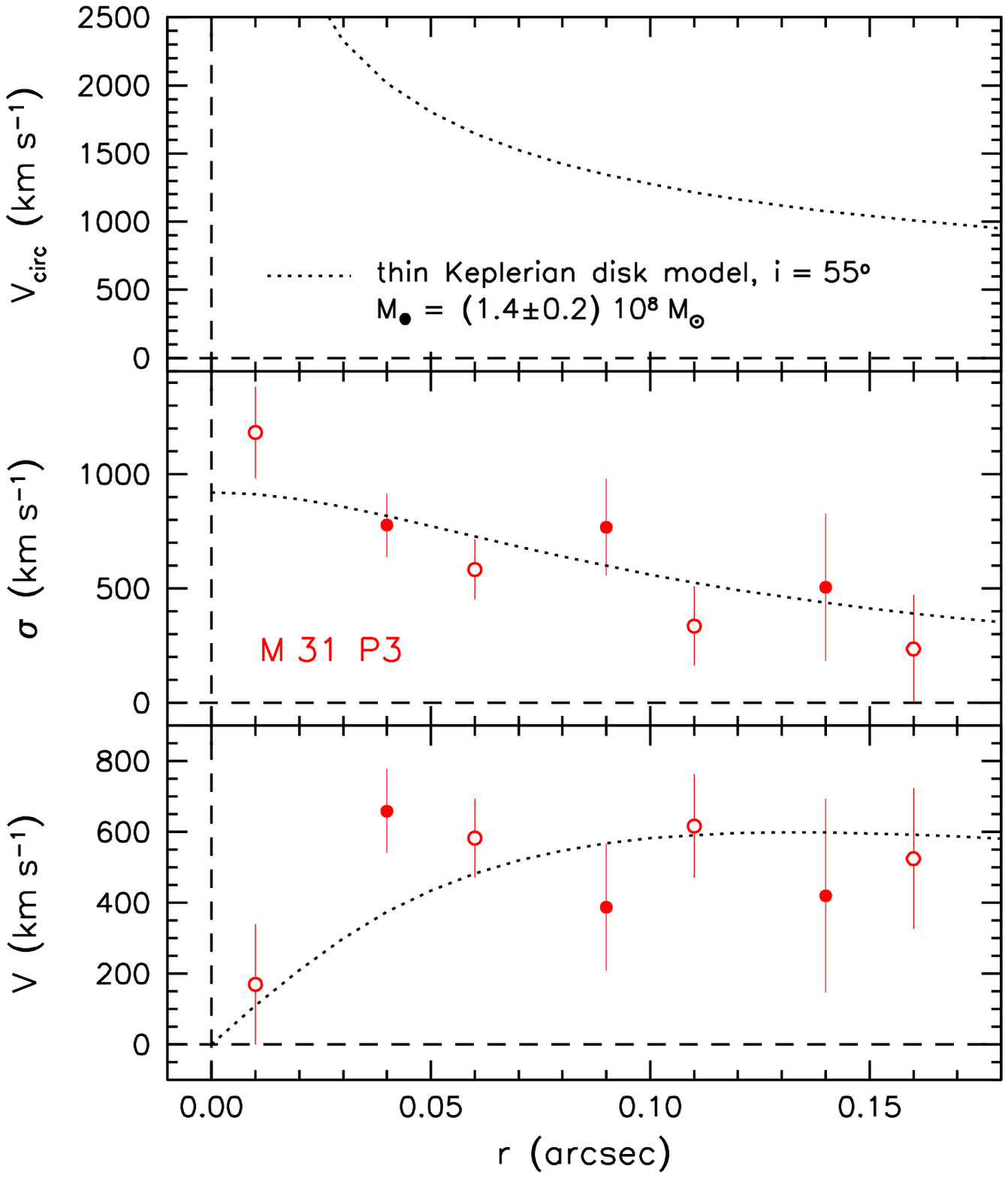,width=8.52cm,angle=0}}
\figcaption[p3_diskmodel_sb2.ps]{The three color images at the top
show the PSF-broadened thin disk model of P3. The images cover
$0\farcs5 \times 0\farcs5$.  Shown from left to right are: (i) P3
surface brightness; intensities range from 0 (black) to 1 (white);
(ii) P3 rotation velocity field with the slit and radial bins 
superposed; the velocity amplitudes range from $-700$ km s$^{-1}$
(black) to $+700$ km s$^{-1}$ (white); (iii) P3 velocity dispersion,
ranging from 150 km s$^{-1}$ (black) to 1000 km s$^{-1}$ (white).
The panels of plotted data points show the P3 radial profiles 
(red) of rotation velocity (bottom) and velocity dispersion (middle), 
folded around P3's center.  Open and closed symbols are from opposite
sides of the center.  The sense of rotation is the same as for the 
eccentric disk P1{\ts}+{\ts}P2.  The top plot shows the best-fitting 
Keplerian circular velocity curve as a dashed line.  It implies a 
black hole mass of $\sim 1.4 \times 10^8$ $M_\odot$.  Convolving the
circular velocity field with the PSF and integrating it over the 
pixel size and slit width yields the model rotation and dispersion 
profiles shown as dotted curves in the bottom and the middle panels.
\label{}}

\vskip 0.5cm


       The PSF-convolved model and the difference between P3 and the
model are illustrated in Figure 9.  We also compare model and P3 with
respect to their isophotal parameters. Figure 10 shows the surface
brightness, ellipticity and position angle profiles of the observed
P3, the model and the PSF-convolved model. Also shown are deconvolved
P3 profiles, which were obtained from 15 iterations with the
Richardson-Lucy deconvolution algorithm implemented in the ESO MIDAS
package. The deconvolved surface brightness profile obtained here
agrees well in shape with the one by Lauer et al. (1998). Figure 10
shows that the model represents P3 reasonably well, especially over
the radius range for which we can analyse the kinematics
(\S\ts6). Surface brightness fluctuations become large at radii beyond
0\farcs25.  Still, the model is adequate out to $\sim$\ts0\farcs4.

      If P3 is an inclined, thin disk, then the observed ellipticity
implies an inclination $i = 55^\circ \pm 2^\circ$.  This is compatible
with the inclination of the eccentric disk P1{\ts}$+${\ts}P2:~Peiris
\& Tremaine (2003) derive $i = 54^\circ$, and Bacon et al.~(2001) get
$i = 55^\circ$.  The model parameters of P3 are summarized in
Table~2.  Whether P3 really is a thin disk can only be checked with
kinematical data.  We discuss these in the next section.

\section{Dynamics of P3}

      Figure 11 shows the rotation velocity and velocity dispersion 
profiles of P3.  Table 3 lists the data.  We used FCQ for the
analysis but did not fit the $h_3$ and $h_4$ Gauss-Hermite parameters
because the S/N of the data is only $\sim 3$ per \AA.  Outside of the 
central pixel, P3 rotates rapidly, with an observed amplitude of 
$573 \pm 61$ km s$^{-1}$ (weighted mean of all points with $|r| >
0\farcs01$).  P3 rotates in the same sense as P1{\ts}$+${\ts}P2. 
The apparent velocity dispersion drops from $\sim$\ts1200 km s$^{-1}$ 
in the central pixel to $<$ 500 km s$^{-1}$ at $r = 0\farcs15$ = 0.55 pc.
These values are consistent with the velocities seen in the extreme 
wings of the line-of-sight velocity distribution of the 
red stars at $r \sim -0\farcs1$ (see Appendix). The kinematic data
securely locate the BH at the center of P3 with an uncertainty of
about 1/3 of a pixel $= 0\farcs02 = 0.07$ pc.

\begin{center}
\textsc{Table 3}

\vskip 0.1cm 

\textsc{Kinematics of P3}

\vskip 0.2cm 

\begin{tabular}{rrrrr}
\tableline
\tableline
\tablevspace {3pt}
     radius   &    V  &   $\Delta$V &  $\sigma$ & $\Delta\sigma$\\   
     arcsec   &   km s$^{-1}$ & km s$^{-1}$ & km s$^{-1}$ & km s$^{-1}$ \\ 
\tablevspace {3pt}
\tableline 
\tablevspace {3pt}
 $-0.16$ & $ 525$ & 197 &  237 & 233 \\
 $-0.11$ & $ 616$ & 144 &  337 & 170 \\
 $-0.06$ & $ 582$ & 111 &  583 & 131 \\
 $-0.01$ & $ 170$ & 169 & 1183 & 200 \\
 $+0.04$ & $-659$ & 117 &  777 & 139 \\
 $+0.09$ & $-387$ & 179 &  769 & 211 \\
 $+0.14$ & $-420$ & 273 &  505 & 322 \\
\tablevspace {3pt}
\tableline
\end{tabular}
\end{center}

\vskip 0.5 cm

      We wish to combine the surface brightness data (Table 2) and the
kinematic data (Table 3) to make dynamical models.  Because the pixel
size, slit width, and PSF are all similar to the size of P3, unresolved 
rotation must contribute to the apparent velocity dispersion. Actually, 
almost all of the light of P3 falls into the slit.  Despite this and 
despite the modest apparent flattening, P3's apparent rotation velocity 
and velocity dispersion are similar. Therefore, it is reasonable to 
expect that P3 is an intrinsically flat object with \hbox{$V \gg \sigma$}.

      For these reasons, we first model P3 as a flat disk with an 
exponential profile and an inclination $i \sim 55^\circ$ (\S\thinspace6.1).
Then (\S\thinspace6.2), we explore more nearly edge-on models in which P3 
has some intrinsic thickness.

\subsection{P3 as a Flat Exponential Disk}

\lineskip=0pt \lineskiplimit=0pt

      We construct a dynamical model in which we assume that P3 is a
flat disk with the parameters in Table~2 and a negligible instrinsic
velocity dispersion.  The BH affects the structure of the galaxy
interior to $r_{\rm cusp} \simeq G M_\bullet / \sigma^2 =
5\farcs6\ts[M_\bullet/(10^8{\ts}M_\odot)]$, where $G$ is the
gravitational constant and $\sigma = 145$ km s$^{-1}$ (Kormendy 1988)
is the velocity dispersion of the bulge just outside the region
affected by the BH.  Since P3 is tiny compared to $r_{\rm cusp}$, the
black hole dominates the gravitational potential.  The distribution of
the stars is completely constrained by the photometry, so the only
free parameter is the BH mass.  To compare the model with the observed
rotation and velocity dispersion profiles, we convolve the Keplerian
velocity field with the PSF and integrate it over the 0\farcs2 slit
width and 0\farcs05 CCD pixels (see Figure 11, top-middle panel).
This is done with small subpixels to obtain smooth profiles of
rotation velocity and velocity dispersion. 

      Figure 11 shows the results.  The observed rotation and dispersion 
profiles (open and closed symbols) are well matched by the model (dotted 
curves).  Estimating the mass of the black hole is straightforward, 
because $M_\bullet$ is the only free parameter.  The best fit gives 
$M_\bullet = (1.4 \pm 0.2) \times 10^8$ M$_\odot$.  The reduced 
$\chi^2_n$ is $\sim$\ts1 (Figure 12). 

      The BH mass derived with the thin disk model does not depend
significantly on inclination over the range allowed by the photometry 
($\pm 2^\circ$).  Changing the inclination away from the best value
increases $V$ and decreases $\sigma$ or vice versa.  Then $\chi^2$
increases slightly, but the shape of the $\chi^2$ distribution as a
function of black hole mass does not change significantly.  We also
varied the scale length of the P3 disk, its total luminosity, and its
position angle on the sky within the errors.  There was no significant
effect on $M_\bullet$.  The total luminosity and mass of P3 are
irrelevant provided that the BH dominates the potential.  The
position angle would have to be changed well beyond its estimated
errors to achieve a visible effect on the velocities.  Changing the
radial scale length redistributes light and makes the rotation and
dispersion profiles flatter or steeper.  Within the errors,
$M_\bullet$ is not affected.

      The circular velocities for the thin disk model are shown in the
top panel of Figure 11.  Future observations that resolve individual
stars should see velocities of 1000 to 2000 km~s$^{-1}$.  Such 
measurements can also test how much the observed velocities and 
dispersions are affected by shot noise due to the small number of stars
in P3.  Checking how close to circular the disk really is will also
be important.

      If P3 is a thin stellar disk, can it be stable? The answer is
yes, as long as its stellar mass is not very much larger than 5200
$M_\odot$.  Even relatively small dispersions will not lead to
significant two-body relaxation.  Using Equation 8-71 in Binney \&
Tremaine (1987), we obtain relaxation times of the order of a Hubble
time.  Moreover, the critical velocity dispersion for local stability
(Toomre 1964) is small, $\sigma_{\rm crit} \ll 1$ km s$^{-1}$.  This
is a consequence of the fact that the potential is dominated by the 
black hole.  That is, the P3 disk is dynamically analogous to Saturn's 
rings rather than to a self-gravitating disk.  Therefore, if earlier
starbursts contributed mass without affecting its present spectrum,
the P3 stellar disk is likely to be locally stable and immune from
two-body relaxation.  And if P3 consists only of young stars, then it
has not had time for significant dynamical evolution.

\subsection{P3 Schwarzschild Models}

\lineskip=0pt \lineskiplimit=0pt

      To investigate the effect on $M_\bullet$ of allowing P3 to have
some thickness in the axial direction $z$ and therefore to be more
nearly edge-on than $i = 55^\circ$, we fitted Schwarzschild (1979) 
models to the photometric and kinematic data.  We used the regularized
maximum entropy method as implemented by Gebhardt et al. (2000a, 2003) 
and by Thomas et al.~(2004).  The program was constrained to reproduce
the observed surface brightness distribution of P3.  We considered 
three inclinations $i = 58^\circ$, $66^\circ$, and $90^\circ$, 
corresponding to intrinsic axial ratios of P3 of $0.26$, $0.44$, 
and $0.57$, respectively.  Black hole masses were varied until the 
kinematic data were reproduced as well as possible, as indicated by
the $\chi^2$ values in Figure 12.

      In the Schwarzschild code, phase space is quantized on a polar 
grid that is not optimized for closed orbits.  It is therefore helpful
if the orbits are not quite closed.  For this reason, we did not use
a point mass for the central dark object but rather used a Plummer 
sphere with a half-mass radius $r_h = 0\farcs01$.  Given the spatial
resolution of the data, this is essentially equivalent to a black hole 
(see Figure 14).

      Models that put significant weight on entropy maximization did 
not fit the kinematics.  They rotated too slowly, because they contained
retrograde orbits.  This is expected, because entropy maximization 
is not appropriate for highly flattened systems with strong rotational 
support.  

      Switching off the entropy maximization (this corresponds to 
a high regularization parameter in Thomas et al.~2004) results 
in better fits.  Figure 12 shows $\chi^2$ values as a function of
inclination and dark mass $M_\bullet$.  We conclude that the lowest
inclination, $i = 58^\circ$, is preferred, by $\Delta\chi^2 \approx
2$ relative to the $i = 66^\circ$ model and with higher significance
relative to the more inclined models.  

      Rotation velocity and velocity dispersion profiles for the 
lowest-$\chi^2$ model at each inclination are shown in Figure 13. 
Reassuringly, the $i = 58^\circ$ Schwarzschild model most nearly 
resembles the $i = 55^\circ$ thin disk model.  The fits then become 
progressively more different -- and less good -- as the models are 
made more edge-on.

      Higher inclinations require higher BH masses.  The 
reason is that, at higher inclinations, line-of-sight integration 
through the nearly edge-on, thick disk includes stars at relatively
large radii that move mostly across, not along, the line of sight.
They reduce the velocity moments and consequently require higher
$M_\bullet$ to match the observed rotation velocities.  The preferred 
black hole mass for the $i = 58^\circ$ and $i = 66^\circ$ Schwarzschild
models is $\sim 2 \times 10^8$ $M_\odot$.  The highest black hole 
mass that is consistent with the data to within $\sim 1 \sigma$ is 
given by the $i = 58^\circ$ Schwarzschild model and is $\sim 2.3 \times
10^8$ $M_\odot$ (Figure 12).  The lowest black hole mass implied by 
the dynamics of P3 is given by the thin disk, $i = 55^\circ$ model and is 
$\sim 1.2 \times 10^8$ $M_\odot$ (\S\thinspace6.1).

      It is instructive to examine the $i = 58^\circ$ Schwarzschild
model in more detail.~Figure 14 shows its velocity moments.~Rotation 
dominates the dynamics; $V_\phi$ is $\lesssim$ 20\thinspace\% smaller 
than the circular velocity.   At radii $r \gtrsim 0\farcs1$, the 
model is approximately isotropic.  To provide the thickness that is 
necessitated by the inclination $i > 55^\circ$, $\sigma_z$ then
increases substantially inward, although it remains smaller than the
rotation velocity.  The difference between the adopted Plummer model 
and a central point mass is small except for the innermost data point. 

\centerline{\null}

\centerline{\null}

\centerline{\null}

\centerline{\null}

\vfill

\vskip 0.3cm
\centerline{\psfig{file=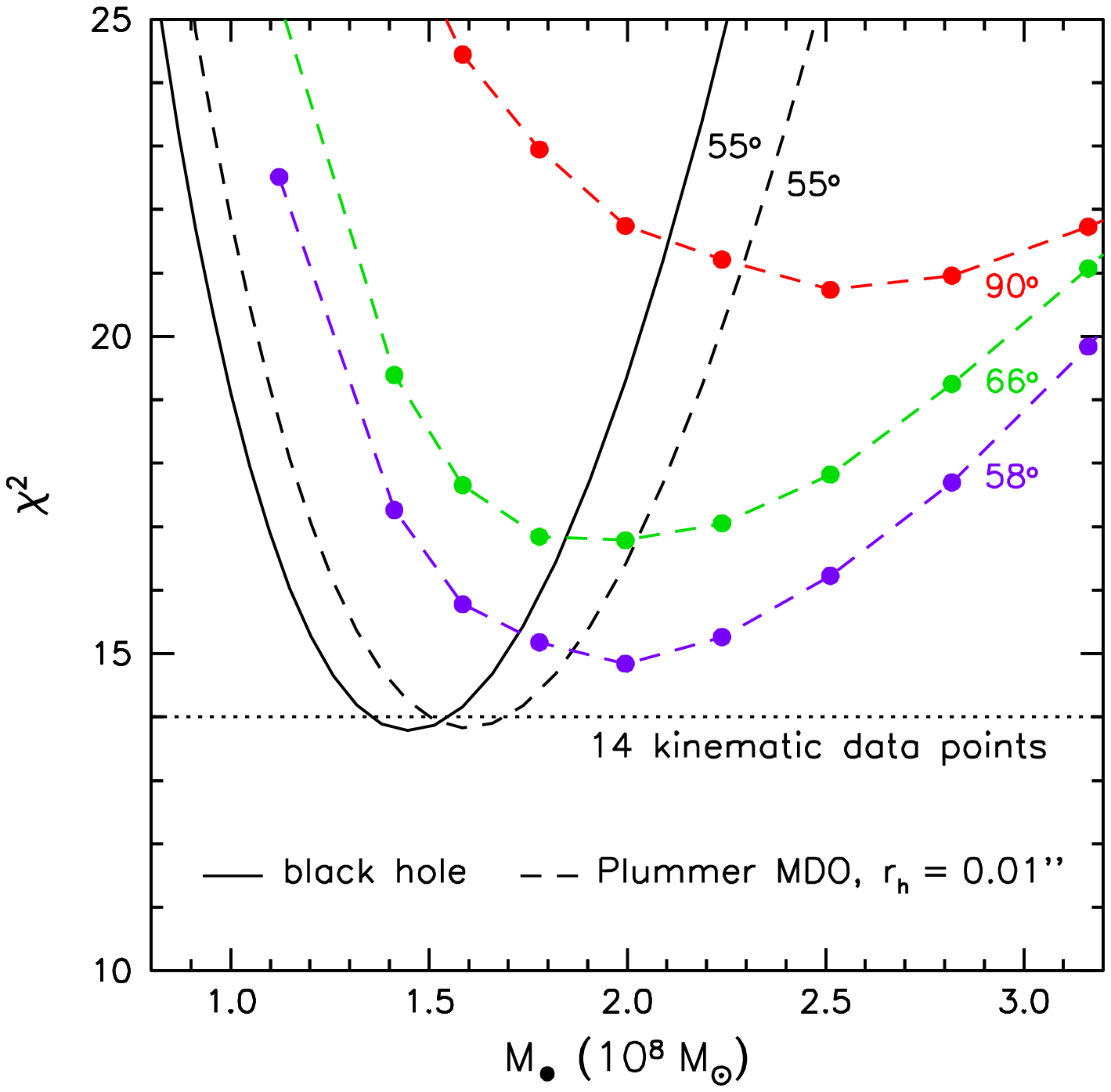,width=8.6cm,angle=0}}
\figcaption[p3_iso_nf.ps]{$\chi^2$ profiles for models of P3. The
black lines refer to the thin disk model with inclination $55^\circ$,
the coloured lines show three Schwarzschild models with inclinations
$58^\circ$, $66^\circ$, and $90^\circ$. The dashed lines assume that
the massive dark object is a Plummer sphere with radius $0\farcs01$,
the full black line assumes a black hole. 
\label{fig12}}

\vfill

\vskip 0.3cm
\centerline{\psfig{file=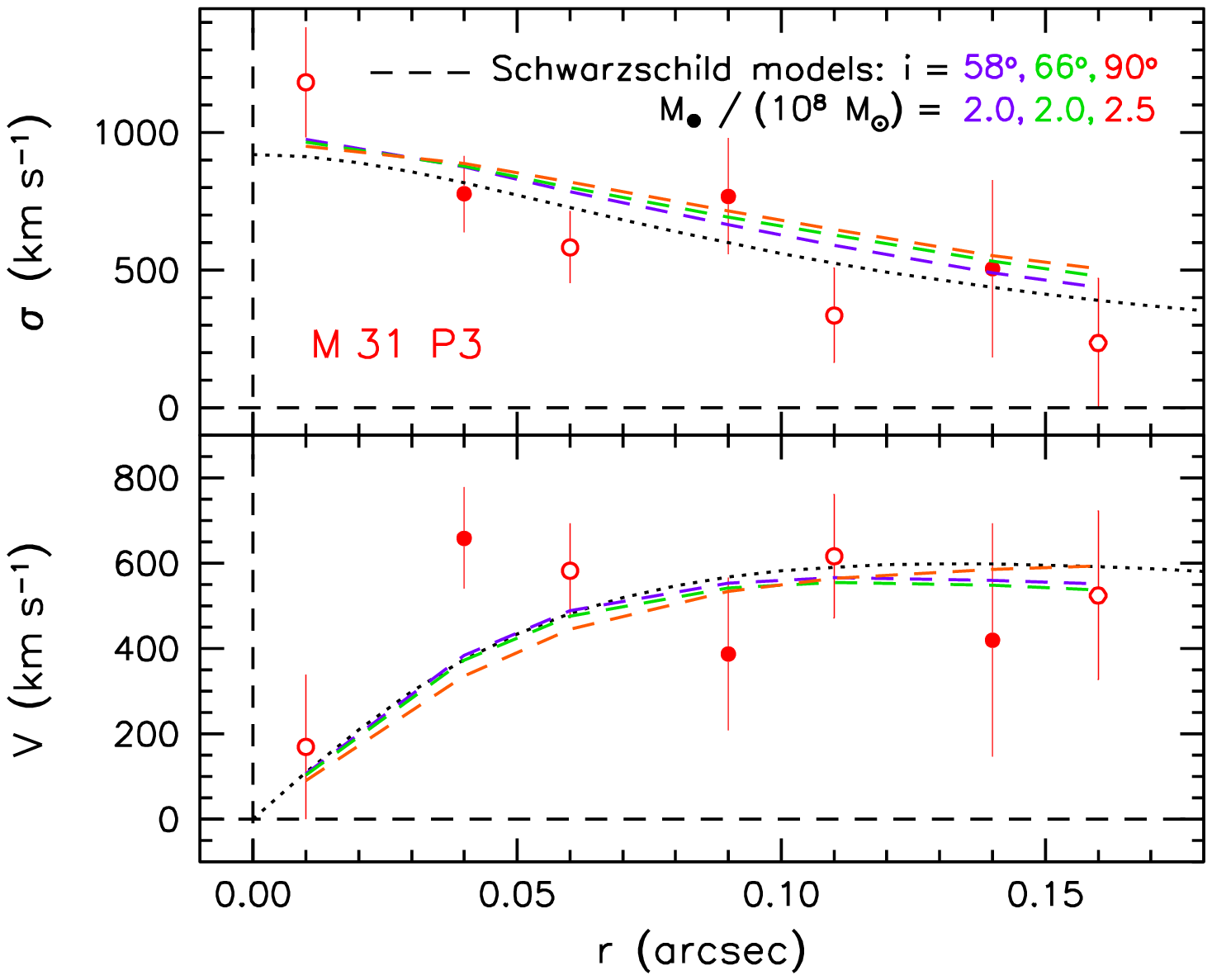,width=8.6cm,angle=0}}
\figcaption[p3_iso_nf.ps]{Rotation velocities and velocity dispersions
of P3 as in Figure 11 (red open and closed symbols). Overplotted as
colored dashed lines are three Schwarschild models with different
inclinations and black hole masses. The thin disk model of Figure 11
is shown as a dotted black line.
\label{fig13}}
\vskip 0.3cm

\vfill

\centerline{\null}

\vfill

\clearpage

\def\bk{\kern -2pt}

\vskip 0.3cm
\centerline{\psfig{file=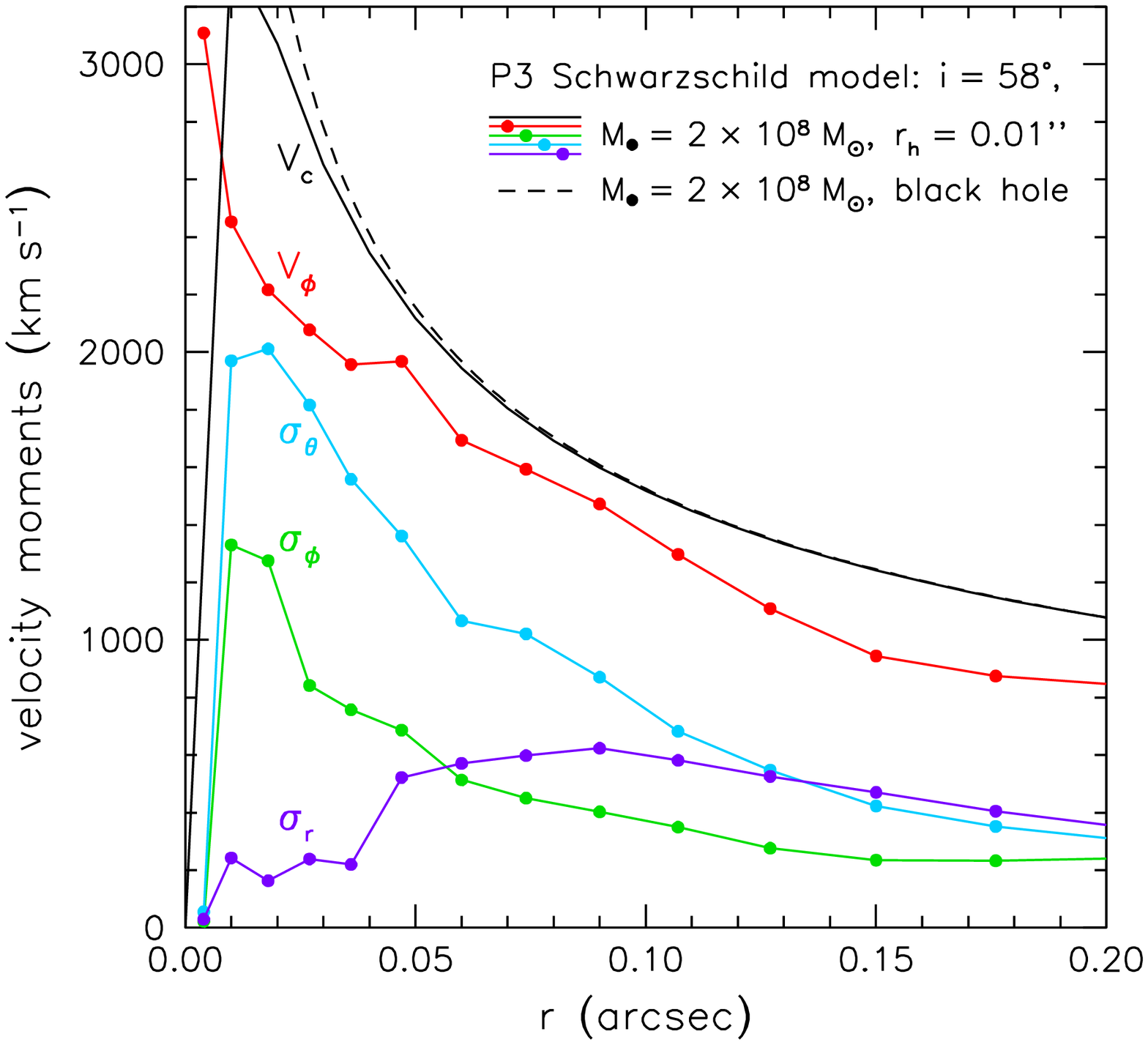,width=8.6cm,angle=0}}
\figcaption[p3_iso_nf.ps]{Major-axis velocity moments of the best
$i= 58^\circ$ Schwarzschild model for P3, corresponding to $M_\bullet 
= 2 \times \bk10^8$ $M_\odot$. The MDO is a Plummer sphere with half-mass
radius $r_h = 0\farcs01$; its circular velocity is shown as a solid
black line.  A BH of the same mass would produce the Keplerian circular
velocities indicated by the dashed curve.
\label{fig14}}
\vskip 0.3cm

\vskip 0.3cm
\centerline{\psfig{file=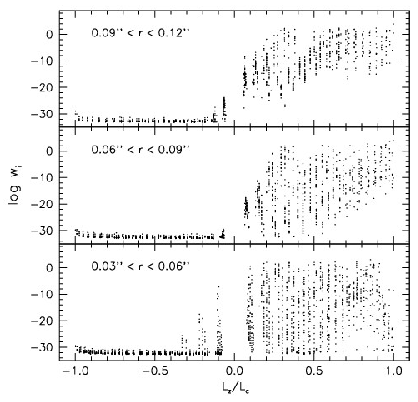,width=8.6cm,angle=0}}
\figcaption[p3_iso_nf.ps]{Orbit structure of the Schwarzschild model
with $i= 58^\circ$, Plummer model dark mass $M_\bullet = 2\times 10^8$
$M_\odot$, and half-mass radius $r_h = 0\farcs01$. For each orbit, the
orbit weight $w_i$ per phase space volume is shown as a function of 
the $z$ component of its angular momentum $L_z$ normalized by the 
angular momentum $L_c$ of the circular orbit that has the same energy.
In this figure only, $r$ is the average of the pericenter and apocenter
radii of the orbit.  Note that at all radii, only prograde orbits are 
significantly populated.
\label{fig15}}
\vskip 0.75cm 

      Figure 15 shows the corresponding orbit structure.  As expected
for a Schwarzschild model that is not too different from the thin disk 
model, retrograde orbits are strongly suppressed.  However, as 
indicated by Figure 14, noncircular orbits get significant weight in
order to produce that axial ratio $b/a \simeq 0.26$.  As expected, 
this happens more near the center than at larger radii.  However, 
nearly circular orbits dominate; otherwise $V_\phi$ would not be almost 
equal to $V_c$ in Figure 14. 

\subsection{Summary: Comparison of P3 and P1{\ts}$+${\ts}P2}

      We conclude that the triple nucleus of M31 is made of two nested,
disk-like systems.  The P1{\ts}$+${\ts}P2 disk is elliptical, has a
radius of about 8 pc, and consists of old, metal-rich stars. If it is
thin, it has an inclination of $\sim 54^\circ$ and a major-axis position
angle of $\sim 43^\circ$ (Peiris \& Tremaine 2003).  The P3 disk is
approximately circular and has a radius of about 0.8 pc.  If it is thin, 
P3 has an inclination $i \sim 55^\circ$ that is the same as that of 
P1{\ts}$+${\ts}P2.  P3's major-axis position angle at $r \leq 0\farcs25$
is $\sim$~$63^\circ \pm 2^\circ$.  That is, the inner P3 disk is slightly 
tilted with respect to the P1{\ts}$+${\ts}P2 disk but is relatively close 
to the kinematical major axis, P.A.$ \sim 56^\circ$, found by Bacon et 
al.~(2001).  At $r > 0\farcs25$, the major axis of P3 twists to $\sim
40^\circ$, essentially the position angle of P1{\ts}$+${\ts}P2.  The
nested disks rotate in the same sense and have almost parallel angular
momentum vectors.

\section{The Mass of the Central Dark Object}

      We have demonstrated that disk-like models for P3 fit both the
photometry and the kinematics of P3 exceedingly well.  This allowed an
estimate of $M_\bullet$ that is independent of all previous
determinations.  Besides black hole mass, only inclination is a
free parameter in the fit to the rotation curve and the dispersion
profile (Figures 11, 13).
   
       Could systematic effects cause additional errors that are not
included in the statistical errors, especially toward low BH masses?
We mentioned in the previous section that some clumpiness in the
distribution of stars could be hidden by PSF blurring and may affect
the measured velocities and dispersions. We carried out a simple check
for this effect by fitting subsamples of data points. E.g., a fit to
just the innermost three points in Figure~11 typically yields BH
masses about 15\% higher, while omitting these three points results in
25\% lower masses. All values obtained in this way fell in the range
allowed by the $\chi^2$-profiles in Figure 12, so this effect does not
seem to be very important.
 
      Some non-circularity of the P3 disk could be hidden as well.  
P3 could contain stars on elongated orbits that have their
pericenters within the range of the kinematic data (0\farcs15) but
apocenters spread out over radii well beyond this radius.  Figure 9
shows faint blue stars that could be such objects.  This would imply
that P3's velocity amplitudes are increased by rotation velocities
that are faster than circular. It is difficult to estimate the size of
this effect, but pericenter velocities of very radial stars can be at
most a factor of 2 larger than pericenter velocities of stars on
nearly circular orbits. Averaging over a set of orbits will reduce this
number considerably. And if we wanted to fully exploit this effect,
many more stars of P3 would have to be found outside of $\sim
0\farcs18$ than inside, which is in contradiction with the
observations. The Schwarzschild models also show that this trick 
does not work well. More radially biased models (obtained with
higher entropy weighting and not shown here) require larger BH masses. 
Finally, if P3 originated in a star-forming gas disk, it
could not contain nearly radial orbits, and we noted above that
subsequent internal evolution of the P3 disk should be slow.

      So, very special circumstances would be required to decrease
$M_\bullet$ below $1 \times 10^8$ $M_\odot$. On the high mass end, the
black hole mass grows with increasing inclination of the model.
However, the $\chi^2$ values become inacceptably large for
inclinations above $\sim 60^\circ$ and, therefore, it is unlikely that
the black hole mass is significantly larger than $\sim 3\times 10^8
M_\odot$. Viable models for P3 are found in the inclination range
$55^\circ < i < 58^\circ$ and in the BH mass range $1.1\times 10^8
M_\odot< M_\bullet< 2.3\times 10^8 M_\odot$. The upper limit takes
into account that the Schwarzschild models were calculated assuming an
MDO with $r_h = 0\farcs01$ and not a BH; the upper limit for a BH is
$\approx 0.2\times 10^8 M_\odot$ lower than for an MDO with $r_h =
0\farcs01$.  The best fit and at the same time lowest black hole mass
of $M_\bullet = 1.4 \times 10^8 M_\odot$ is obtained for the thin disk
model. This model is also preferred on astrophysical grounds, if P3
formed out of a thin gaseous disk.  {\it Therefore, our best estimate
for the mass of the supermassive black hole in M{\thinspace}31 is
$M_\bullet = 1.4^{+0.9}_{-0.3} \times 10^8 M_\odot$.}

      How does this compare with previous results?  The black hole
mass has now been estimated by five, largely independent techniques,
(i) standard dynamical modeling that ignores asymmetries, (ii) the KB
center-of-mass argument that depends on the asymmetry of
P1{\ts}$+${\ts}P2, (iii) the Peiris \& Tremaine nuclear disk model
that explains the asymmetry of P1{\ts}$+${\ts}P2, (iv) full dynamical
modeling that takes into account the self-gravity of the
P1{\ts}$+${\ts}P2 disk (Salow \& Statler 2004), and (v) dynamical
modeling of the blue nucleus P3, which is independent of
P1{\ts}$+${\ts}P2.  The good news is that all methods require a dark
mass with high significance.  The bad news is that some of the results
differ by more than two standard errors.  In particular, the
disagreement between the KB center-of-mass argument and the P3 models
presented here is a concern.  We therefore revisit the KB derivation
in the subsection below.  The models that best fit both the photometry
and the kinematics -- the Peiris \& Tremaine (2003) eccentric disk
model of P1{\ts}$+${\ts}P2 and our thin disk model of P3 -- agree
within the errors and favor a high black hole mass of $M_\bullet \sim
1 \times 10^8$ $M_\odot$.  We also note that a higher black hole mass
can be accommodated more easily in almost all models than a lower
black hole mass.

      The mass of the M{\ts}31 BH derived here is a factor of $\sim$
2.5 above the ridge line of the correlation between $M_\bullet$ and bulge
velocity dispersion $\sigma_{\rm bulge}$ (Ferrarese \& Merritt 2000;
Gebhardt et al.~2000b).  Using the Tremaine et al.~(2002) derivation,
$$ \log\left(\frac{M_\bullet}{M_\odot}\right)\; =\; 8.13\; +\; 4.02\,
\log\left(\frac{\sigma_{\rm bulge}}{200~\rm{km\ s}^{-1}}\right)~,$$
$\sigma_{\rm bulge} \simeq 160$ km s$^{-1}$ implies that $M_\bullet
\simeq 5.5 \times 10^7$ $M_\odot$.  We derive $M_\bullet =
1.4^{+0.9}_{-0.3} \times 10^8$ $M_\odot$. Tremaine et al. (2002)
already found significant scatter in the $M_\bullet-\sigma_{\rm
bulge}$ relation at low masses. With the increased BH mass for M~31,
scatter has become even more prominent. Considering, in addition to
M~31, only the two closest other supermassive BHs, i.e. M~32 and the
Galaxy, we get the following.  M{\ts}32 has $\sigma_{\rm bulge} \sim
75$ km s$^{-1}$, a predicted $M_\bullet = 2.6 \times 10^6$ $M_\odot$
and an observed $M_\bullet = (2.9 \pm 0.6) \times 10^6 M_\odot$
(Verolme et al.~2002 corrected to distance 0.81 Mpc from Tonry et
al.~2001).  Our Galaxy has $\sigma_{\rm bulge} \sim 103$ km s$^{-1}$,
a predicted $M_\bullet = 9.4 \times 10^6$ $M_\odot$ and an observed
$M_\bullet = (3.7 \pm 0.2) \times 10^6$ $M_\odot$ (Ghez et al.~2004).
So M{\ts}31, M{\ts}32, and our Galaxy have black hole masses that are
2.5 times larger than, consistent with, and 3 times smaller than the
ridge line of the $M_\bullet-\sigma_{\rm bulge}$ relation,
respectively.  This is strong indication for significant intrinsic
scatter in the $M_\bullet-\sigma_{\rm bulge}$ relation, at least at
the low-mass end.

\vskip 0.3cm

\subsection{Black Hole Mass From The Center-of-Mass Argument:\\ 
            Kormendy \& Bender (1999) Revisited}

      KB estimated that the center of P3 (blue dot in our Figure~16) is
offset from the bulge center (horizontal dashed line) by about 0\farcs06. 
They then assumed that the central dark object is in P3 and estimated its
mass based on the assumption that the combined system, 
BH{\ts}$+${\ts}P1{\ts}$+${\ts}P2{\ts}$+${\ts}P3, is in dynamical equilibrium.  
That is, they assumed that the center of mass of 
BH{\ts}$+${\ts}P1{\ts}$+${\ts}P2{\ts}$+${\ts}P3 is at the center of the bulge.  
Then $M_\bullet$ is inversely proportional to its distance from the bulge 
center and, if the mass of P3 is negligible, $M_\bullet$ is proportional
to the mass of P1{\ts}$+${\ts}P2.  The latter was given by
the light distribution and the measured mass-to-light ratio $M/L_V =
5.7$.  The resulting black hole mass was $M_\bullet \simeq (3.3 \pm
1.5) \times 10^7$ M$_\odot$.  This is at the low end of the range of
published values and a factor~of~$\sim$\ts4 smaller than the value derived
here.  As the stellar $M/L_V$ can hardly be a factor 4 larger, two
explanations are possible for the discrepant BH masses:~(i) P3 and the 
BH are a factor of 4 closer to the bulge center than KB derived, or (ii)
the BH{\ts}$+${\ts}P1{\ts}$+${\ts}P2{\ts}$+${\ts}P3 system is not in 
equilibrium with respect to the bulge center.

      We believe that the observations can be consistent with
equilibrium and that the distance of P3 to the bulge center was
overestimated by KB for the following reasons.


\vskip 0.3cm \psfig{file=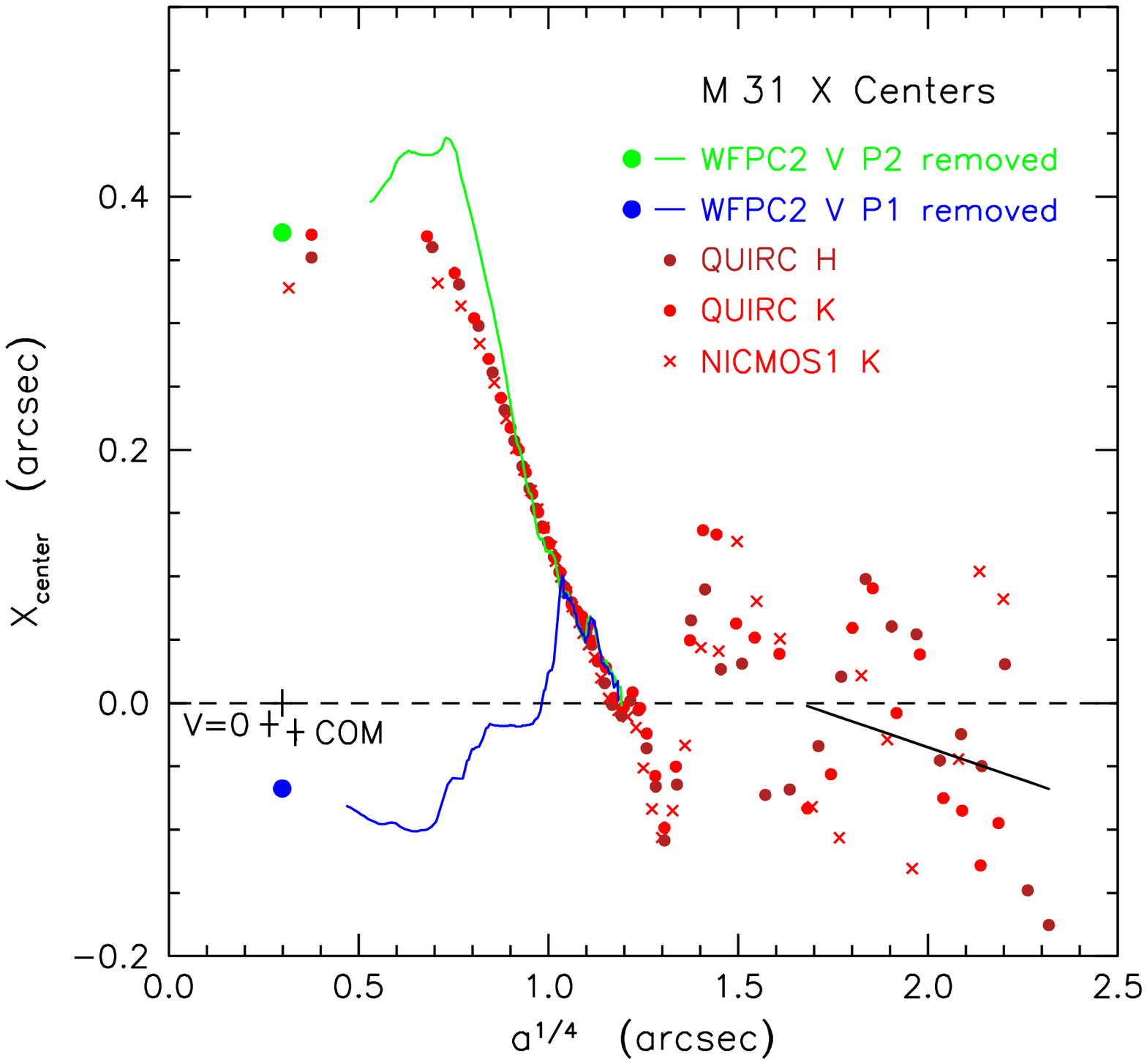,width=7.3cm,angle=0}
\includegraphics{white.ps}
\figcaption[progentors.ps]{From KB, isophote center coordinates $X$
along the line joining P1 and P2 as a function of isophote major-axis
radius $a$.  A $V$-band HST WFPC2 image was measured twice, once
masking out P2 (green) to measure the convergence of the P1 isophotes
on the center of P1 (green dot) and once masking out P1 (blue) to measure
the convergence of the P2{\ts}$+${\ts}P3 isophotes on the center of P3
(blue dot).  The brown and red points are measurements of individual 
isophotes in $H$- and $K^{\ts\prime}$-band images. The positions of the 
velocity center and of the center of mass of the BH and nucleus 
(if $M_\bullet = 3 \times 10^7$ $M_\odot$), each with error bars, are 
shown by the symbols labeled ``V=0'' and ``COM''.  The dashed line at 
$X = 0$ marks the center position of the bulge that was adopted by KB.
It was estimated by averaging all isophote center coordinates at 
$2\farcs9 < a < 25\farcs0$ ($1.3 < a^{1/4} < 2.24$).  However, if 
$M_\bullet = 1.4 \times 10^8$ $M_\odot$, then the BH's radius of influence
is $r_{\rm cusp} \simeq 7\farcs2$.  Therefore KB estimated the bulge
center position partly from isophotes that are at $a < r_{\rm
cusp}$.  Within this radius, the BH dominates the potential and
isophotes do not need to be concentric to be in equilibrium (witness
the eccentric disk).  Since we now believe that the BH mass is large,
we should derive the bulge center from correspondingly larger radii. 
The solid line is a least-squares fit to the bulge $X_{\rm center}$ values
at $a > r_{\rm cusp}$.  It shows that the isophote centers at the largest
radii in the figure are approximately at the $X$ coordinate of P3. 
Therefore the BH is close to the luminosity-weighted center of the bulge.
\label{}}
\vskip 0.7cm


Figure 16 revisits the center of mass argument.  It is reproduced from
KB and shows their estimate of the position of the center of the bulge
as the dashed line at $X = 0$.  The isophote center coordinate $X$ is
measured along the line joining P1 and P2. A conclusion about the
position of the center of the bulge depends on the radius range chosen
in which to average isophote $X$ values.  KB calculated the average at
$2\farcs9 < a < 25\farcs0$. A larger radius range was not possible
because of the small size of the high-resolution images used.  If
$M_\bullet \sim 3 \times 10^7$ $M_\odot$, then the above radius range is
no problem -- it is beyond the radii affected by the BH.  However, if
$M_\bullet$ is as big as $1.4 \times 10^8$ $M_\odot$, then $r_{\rm
cusp} \simeq 7\farcs2$ and it is necessary to calculate the mean bulge
$X$ at larger radii\footnote{What we should expect at $r < r_{\rm
cusp}$ is not clear.  Because the potential is dominated by the BH,
asymmetries like those of the P1 -- P2 eccentric disk are possible and
isophotes do not need to be concentric to represent an equilibrium
configuration.}.

      If we calculate the bulge center outside of $r_{\rm cusp} \simeq
7\farcs2$, we obtain a mean $X$ position of $-0\farcs033$ in Figure
16, i.{\ts}e.,~half way between the KB bulge center and P3. Note
that, unlike KB, we do not limit the averaging to points with $a <
25\farcs0$ but now also include two further points that we extracted
from the QUIRC $H$-band image beyond this radius. In addition, we omit all
center coordinates with errors larger than $0\farcs2$.  

      A least-squares fit to the points with $a > 7\farcs2$ gives the
short black line in Figure 16.  It shows that the bulge isophote
centers drift with increasing radius toward the $X$ position of P3.
So the luminosity-weighted center of the bulge is close to P3. 

      This discussion suggests that the determination of the bulge center
is less reliable than KB assumed. There are three reasons: (i) the BH sphere 
of influence is much larger than KB assumed, (ii) the bulge isophote centers
drift toward P3 with increasing radius beyond $a$ = 2\farcs9, and 
(iii) the isophote centers oscillate -- or at least, fluctuate -- with
radius because of dust or surface brightness fluctuations or perhaps a
physical effect that we have not identified.  If the BH is much more
massive than P1{\ts}$+${\ts}P2, then it is so close to the center of mass
that $M_\bullet$ cannot be determined accurately from the COM argument. 

      The important conclusion, however, is that the observations of
M{\ts}31 are consistent with dynamical equilibrium and with a large BH mass
of $M_\bullet = 1.4 \times 10^8$ $M_\odot$.

\vskip 0.3cm

\section{Astrophysical Constraints on a Massive Dark Object Made of Dark Stars}

      Central dark masses are detected dynamically in 38 galaxies (see
Kormendy \& Gebhardt{\ts}2001;{\ts}Kormendy{\ts}2004~for~reviews).
They are commonly assumed to be supermassive BHs, although clusters of
dark stars are consistent with the dynamics in most galaxies.
Justifying this assumption, many authors cite the implausibility of
producing so many stellar remnants -- often 100 times the mass in
visible stars -- in the small volume defined by the PSF in which the
dark mass must lie.  Another argument is the consistency of the dark
masses with energy requirements for BHs to power active galactic
nuclei. More rigorous arguments against dark clusters are available
for two galaxies, NGC 4258 and our own Galaxy (Maoz 1995, 1998; Genzel
et al.~1998; Sch\"odel et al.~2002, 2003; Ghez et al.~2004). Clusters
of failed stars are not viable because brown dwarfs collide on short
timescales and either evaporate, or, merge and become visible stars.
Clusters of dead stars are not viable because their two-body
relaxation times are so short that they evaporate.  In NGC 4258, the
timescales associated with these processes are at least as short as
$10^{8.5}$ yr.  In our own Galaxy, they are remarkably short indeed,
$\sim$ $10^4$ yr.  Even balls of neutrinos with cosmologically
allowable neutrino masses are excluded in our Galaxy.  The BH cases in
NGC 4258 and in our Galaxy are now very strong and are taken as
indications that dynamically detected central dark masses in other
galaxies are BHs, too.

      However, a great deal is at stake.  It would be very important if 
astrophysical arguments ruled out BH alternatives in more than two galaxies.
M{\ts}32 has been the next best case (van der Marel et al.~1997, 1998), but 
Maoz (1998, Figure 1) shows that a white dwarf cluster could survive for 
$\sim 10^{11}$ yr.

      Applying our results on the dynamics of P3, M{\ts}31 becomes the third
galaxy in which dark star cluster alternatives to a BH can be excluded. For the
most conservative estimate that $M_\bullet \simeq 3 \times 10^7$ $M_\odot$, 
the arguments are discussed in Kormendy, Bender, \& Bower (2002) and in 
Kormendy (2004).  Here we update these arguments to the best kinematic fits and
resulting BH masses implied by \S\S6 and 7.  More detail is given in Kormendy
et al.~(2005).  

\subsection{Limits on the Size of a Dark Cluster Alternative to a BH}

      Figure 17 derives our adopted limit on the size of any dark
cluster alternative to a BH.  It shows $\chi^2$ countours for fits to
the rotation and dispersion profiles of P3 using the thin disk model
of Figure 11 that gave the lowest black hole mass.  As in Maoz (1995,
1998), we assume that the dark object is a Plummer sphere, i.{\ts}e.,
a reasonably realistic dynamical model with a very steep outer
profile.  We wish to use a relatively truncated mass distribution --
one that is not excessively core-halo -- because we need to fit the rapid
rise in $V(r)$ and the corresponding drop in $\sigma(r)$ (Figure 13)
with a distributed object; dark mass that is at several times the half-mass
radius $r_h$ hurts rather than helps us to do this.

We need to know how large $r_h$ can be and still allow an adequate fit
to the kinematics.  As $r_h$ is increased, the inner rotation curve
drops, and it gets more difficult to fit the high central $\sigma$ and
especially the rapid central rise in $V(r)$.  To compensate, an
adequate fit requires that we increase $M_\bullet$.  Therefore $r_h$
and $M_\bullet$ are coupled (Figure 17).  How extended can the dark
object be?  We adopt the parameters at the upper-right extreme of the
68\ts\% $\chi^2$ contour: $r_h = 0\farcs031 = 0.113$ pc and $M_\bullet
= 2.15 \times 10^8$ $M_\odot$.  Note that choosing values
corresponding to, e.g., the 90\ts\% $\chi^2$ contour (or a larger one)
does not significantly weaken the arguments against a dark cluster
presented below as with increasing $r_h$ the required dark cluster
mass increases as well.


\vskip 0.3cm
\psfig{file=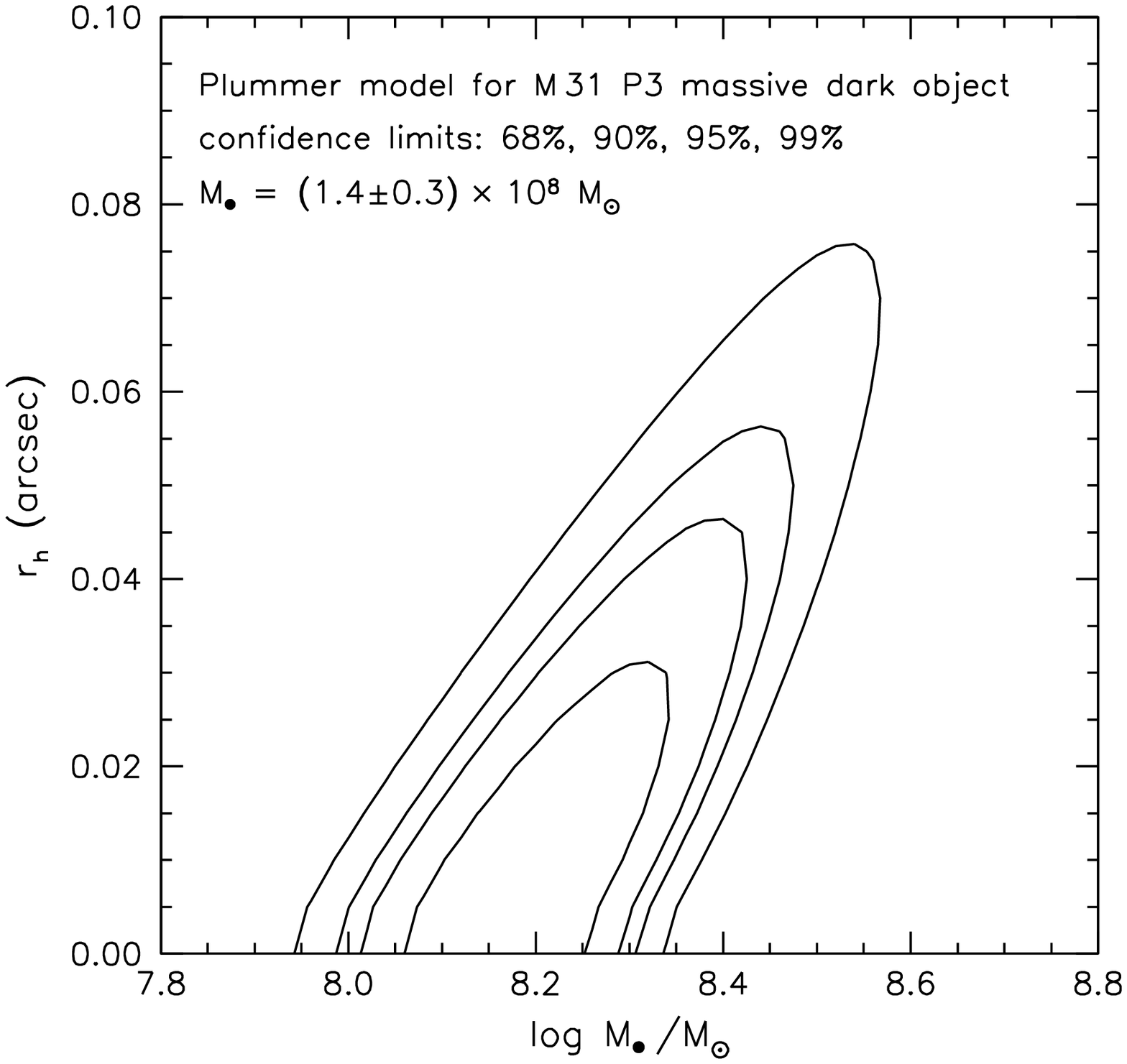,width=8.4cm,angle=0}
\figcaption[]{
Contours of $\chi^2$ for fits to the P3 kinematics of Plummer spheres with
half-mass radii $r_h$ and total masses $M_\bullet$.  The P3 model is a flat 
disk, as in Figure 11.  Our adopted constraint on the fluffiness of the
dark mass is the upper-right extreme of the 68\ts\% $\chi^2$ contour,
i.{\ts}e., $r_h = 0\farcs031 = 0.113$ pc and $M_\bullet = 2.15 \times
10^8$ $M_\odot$.
\label{fig17}}
\vskip 1.2cm


\vskip -30pt

\centerline{\null}

\subsection{Arguments Against a Dark Cluster}

      The half-mass radius $r_h = 0.113$ pc is the same as the radius of the
Ring Nebula (Cox 1999), a typical planetary nebula.  We are considering a
situation in which this volume contains $10^8$~$M_\odot$ of brown dwarf stars 
or stellar remnants.  The mean density inside $r_h$ is $\rho_h = 1.8 \times
10^{10}$ $M_\odot$ pc$^{-3}$, and the density at $r_h$ is $\rho(r_h) = 6.5
\times 10^9$ $M_\odot$ pc$^{-3}$. This is $\sim 10$ times larger than the
largest stellar mass density observed in any galaxy, $7 \times 10^8$ $M_\odot$ 
pc$^{-3}$ at $r = 0\farcs1 = 0.004$ pc in the stellar cusp around Sgr A$^*$ in
our Galaxy (Genzel et al.~2003).  However, only about 300 $M_\odot$ of stars 
are inside the above radius (Genzel et al.~2003).  
Not surprisingly, a dark cluster as extreme as the one that we require
to explain the kinematics of P3 gets into trouble.

\subsubsection{Brown Dwarfs Collide And Destroy Themselves}

      It is easiest to eliminate brown dwarfs. They collide with each
other so violently that they get converted back into gas.  Figure 18
shows the timescale on which every typical brown dwarf collides with
another brown dwarf.  As in Maoz (1995, 1998), the zero-temperature
brown dwarf radius is taken from Zapolsky \& Salpeter (1969) and from
Stevenson (1991), and the calculation is an average interior to $r_h$.
Typical collision velocities at $r_h$ are $\sim 2500$ km s$^{-1}$;
this is fast enough compared to the surface escape velocity ($\sim
600$ km s$^{-1}$ for a 0.08 $M_\odot$ star and smaller for lower mass
brown dwarfs) that the brown dwarfs would get destroyed -- i.{\ts}e.,
converted back into gas.  Brown dwarfs are strongly excluded.


\vskip 0.5cm
\psfig{file=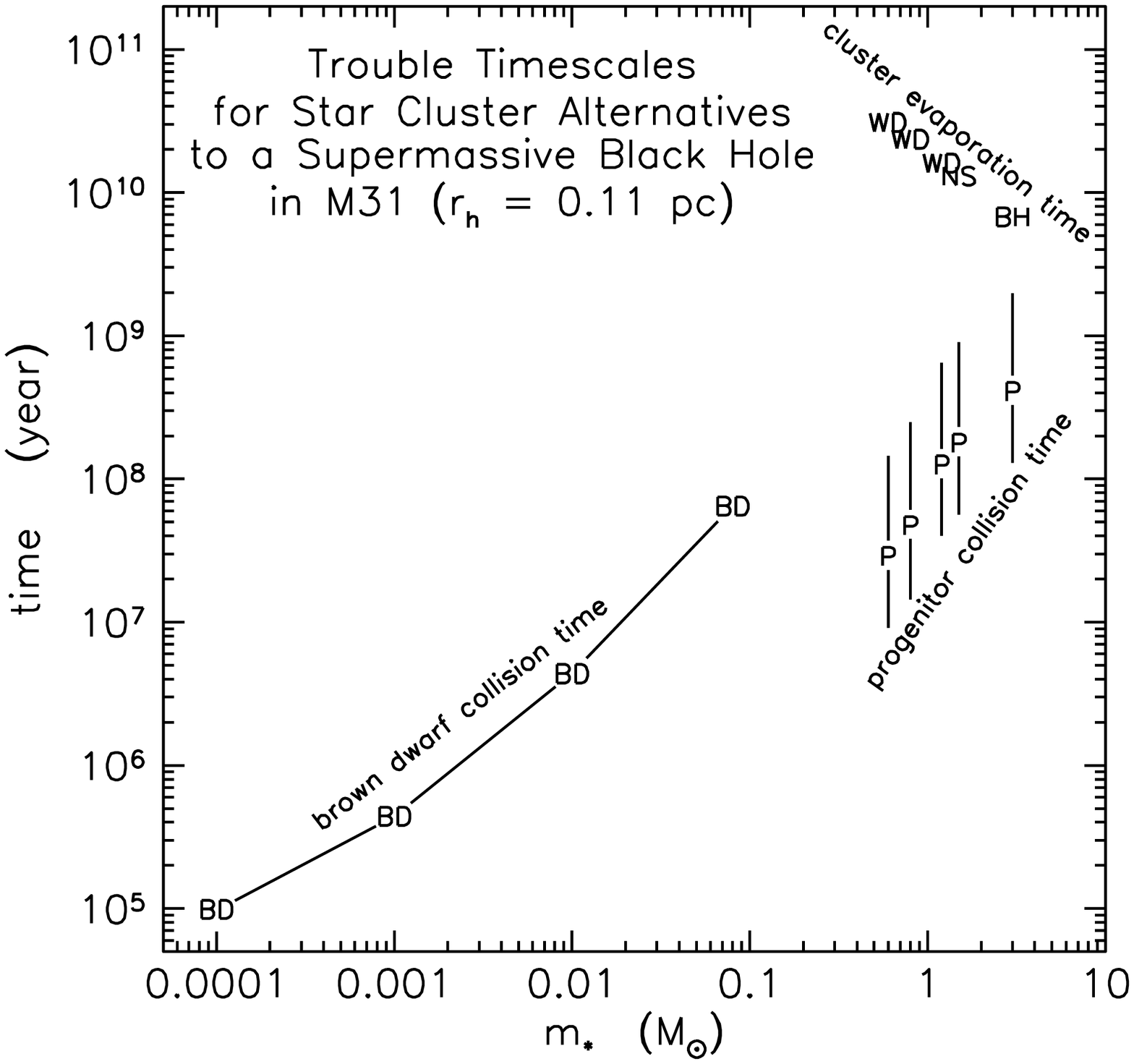,width=4.5cm,angle=0}
\figcaption[progentors.ps]{
Timescales on which dark cluster alternatives to a BH get into trouble in
M{\thinspace}31.  The dark star mass is $m_*$. For a cluster made of brown
dwarfs (BD), the left curve shows the timescales on which every typical star
suffers a physical collision with another star.  Points WD, NS, and BH show the 
times in which dark clusters made of white dwarfs, neutron stars, or
stellar-mass black holes would evaporate.  Points P with ``error bars'' are
the timescales on which every typical progenitor star collides with another 
progenitor at the radius in the Plummer model dark cluster which contains
one-quarter of the total mass.  The letter P is for the time when the dark
cluster is three-quarters assembled; the ``error bars'' end at the collision
times when the cluster is half assembled (top) and fully assembled (bottom).
\label{fig16}}


\subsubsection{Intermediate-Mass White Dwarfs Collide and Make 
               Type Ia Supernovae}

\lineskip=0pt \lineskiplimit=-1pt
\def\bb{\kern -2pt}

      Relatively short collision times provide an argument against 
intermediate-mass white dwarfs.  For 0.8 $M_\odot$ \lapprox \ts$m_*$ \lapprox
\ts1.2 $M_\odot$, collision times at the quarter-mass radius $r_{1/4}$ are
(4{\ts\ts}to{\ts}7)\ts$\times$\ts\bb$10^9$ yr. Given the implied numbers of
white dwarfs interior to this radius and the fact that the collision time would
be shorter at smaller radii, collisions should happen more often than every 50
to 150 yr.  Each collision would bring the remnant well above the Chandrasekhar
limit.  Presumably Type Ia supernovae would result.  Near maximum brightness, 
they would be visible to the naked eye.  The fact that no such supernovae have
been observed in M{\ts}31 might barely be consistent with the above rates, but
if intermediate-mass white dwarfs in similar dark clusters are the explanation
for other galaxies's central dark objects, the resulting supernovae would 
easily have been seen.  Intermediate-mass white dwarfs are implausible.  


      In addition, the supernova ejecta would be lost to the cluster.  The 
above collision times imply that most of the mass inside $r_{1/4}$ and a
significant fraction of the mass inside $r_h$ would be lost in a few billion
years.  For the dark cluster to have its present mass, it would have had to be
more massive in the past.  All problems involving collision rates would get
more severe.

\parindent=8pt

     White dwarfs with masses less than half of the Chandrasekhar
limit will turn out to be excluded because their progenitors would be
destroyed and converted into gas, or, if they succeed to merge, become progenitors of
intermediate-mass white dwarfs or still heavier remnants (\S\ts8.2.4).

\parindent=10pt

      White dwarfs with masses near the Chandrasekhar limit are small.
Their collision times are long.  For these objects, we need stronger arguments.
These arguments will militate against intermediate-mass white dwarfs, also.

\subsubsection{Dark Cluster Formation Scenario:\\
               Let's Imagine Six Impossible Things Before Breakfast}

      Heavy remnants are too small to collide.  Instead, relaxation
gives positive energies to a steady trickle of stars that are lost to the
system.  In $\la$ 300 half-mass relaxation times, the cluster evaporates. 
Figure 18 shows the evaporation times for dark clusters made of 
0.6 M$_\odot$ white dwarfs, 
0.8 M$_\odot$ white dwarfs, 
1.2 M$_\odot$ white dwarfs, 
1.5 M$_\odot$ neutron stars, and 
3 M$_\odot$ black holes (left to right, symbols WD, WD, WD, NS, and BH). Unlike
the case in NGC 4258 and in our Galaxy (Maoz 1995, 1998), these evaporation
times are not implausibly short except for $m_*$ \gapprox \ts10 $M_\odot$ BHs.
So, for most remnants, we need stronger arguments.

      Fortunately, we can add new arguments.  They depend only on canonical,
well understood stellar evolution and on simple stellar dynamics.  A dark 
cluster made of stellar remnants is viable only if its progenitor stars can
safely live their lives and deliver their remnants at suitable radii.  The
properties of the dark cluster constrain how it can form.  We describe the
most benign formation scenario in this subsection.  It requires fine-tuning of
the star formation in ways that we do not know are possible.  However, we will
not base our arguments against the resulting dark clusters on these problems,
because we do not understand star formation well enough.  But main sequence
stars are well understood, and we know progenitor star masses well enough for
the present purposes.  It turns out that progenitor stars get into trouble
because they must be so close together that they collide.  The consequences 
are untenable, as discussed in the following subsections.

      Finding a plausible formation scenario is comparable to imagining
six impossible things before breakfast.~The argument is summarized as follows.
The progenitor cluster must be as small as the dark cluster, because dynamical
friction is too slow to deliver remnants from much larger radii.  From Lauer
et al.~(1998), the density of P2 at $r \sim 0.1$ to 0.2 pc is $\sim$ $10^6$
$M_\odot$ pc$^{-3}$.  Then the characteristic time for dynamical friction
(Binney \& Tremaine 1987, Equation 7-18) to change velocities $v \sim 10^3$
km s$^{-1}$ is $v/(dv/dt) \sim 10^{12}$ yr for 10 $M_\odot$ stars.  This drives
us to imagine the following impossible things:

      1 -- Let's form progenitor stars with a density distribution proportional
to that of the dark cluster; i.{\ts}e., a Plummer sphere with half-light radius
$r_h = 0.113$ pc.

      2 -- We get into less trouble with collisions if fewer progenitors are
resident at one time.  Therefore the safest strategy is to form stars at a
constant rate during the formation time of (say) $10^{10}$~yr.  This is not
the obvious strategy in a hierarchically clustering Universe; it is more natural
to postulate episodic formation by more vigorous events that are connected 
with major mergers.  But shortening the formation time increases the number
of progenitors that must be resident at the same time, and this greatly
increases difficulties with stellar collisions.

      3 -- We assume that all progenitor stars have the same mass.  In
particular, we cannot allow a Salpeter (1955) mass function, because we
cannot tolerate any significant numbers of dwarf stars with lifetimes
long enough so that the stars or their white dwarf remnants remain 
visible today.

      4 -- We assume that sufficient gas for star formation is always
present.  Some gas could come from mass lost by evolving stars,
but some gas must be added continuously to make the cluster grow.
We assume that stars can form despite any energy feedback from massive
or evolved stars.

      5 -- We do not worry about the fact that the young cluster is easy
to unbind gravitationally by the mass loss from evolved stars.  This is
a difficult problem.  Progenitors outmass their remnants by factors of 
at least a few (for low-mass stars) or $\sim 10$ (for high-mass stars).
During the first stellar generations, the progenitors outmass the remnants.
Since they lose most of their mass during the course of stellar evolution, it
is easy to reduce the total mass of the cluster substantially when stars die.  
Impulsive loss of more than half of the mass (say, if the star formation
happened in a coeval starburst) unbinds the cluster.  Slower mass loss
fights the formation process by expanding the cluster.  We ignore all of
these difficulties and assume that the cluster can safely evolve beyond
the fragile initial stage when the mass of progenitors present at one time
is significant.  

      6 -- We assume that the only evolution in $r_h$ is that
resulting from a gradual increase of the cluster mass.  Then $r_h \propto
M_\bullet^{-1}$.

      Using the above assumptions, we calculate the evolution~of the cluster
for various combinations of progenitors and their remnants. Progenitor masses
are from Iben, Tutukov, \& Yungelson (1996) for white dwarfs and from Brown \& 
Bethe (1994) for black holes.  The progenitor clusters get into the
following trouble.

\subsubsection{If M{\ts}31 Is Typical, Then Progenitor Clusters Are Too Bright}

      The above progenitor clusters have absolute magnitudes ranging
from $M_V \simeq -16.3$ to $-17.5$ for the duration of their formation.  These
absolute magnitudes are almost independent of progenitor star mass; higher-mass
progenitors are much more luminous, but they live much less long, so far
fewer are present at one time.  Nuclei as bright as the above could not be
hidden in nearby -- or even moderately distant -- galaxies.  They are rare
(e.{\ts}g., Lauer et al.~1996, 2004).  It is unreasonable to assume that dark
cluster formation lasted for $\sim 10^{10}$ yr in every bulge and then stopped
recently in all galaxies. 

      If the formation of the dark cluster took $\ll 10^{10}$ yr, then
the progenitor clusters are brighter but it is easier to hide them at
large redshifts.  But then all problems that involve stellar
collisions get much worse (see below).

      This problem applies to all types of stellar remnants.

\subsubsection{Dynamical Friction Deposits Remnants At Small Radii}
      
      As noted above, progenitor stars are much more massive than the remnants
of previous generations that already make up the dark cluster.  The
dynamical friction of the progenitors against the remnants makes the 
progenitors sink quickly to small enough radii so that the progenitor 
cluster becomes self-gravitating.  Two problems result.  Progenitor
collision times get shorter; these are discussed further below.  Second,
remnants are deposited at small radii, inconsistent with the density 
distribution that we are trying to construct.  As heavy stars sink, remnants
are lifted to higher radii; the effect is not large for one generation of
progenitors, but it adds up by the time the cluster is finished.  The result
is a dark cluster that is much more core-halo in structure than a Plummer
model.  That is, it is inherently impossible to make a dark cluster that
is as centrally concentrated as a Plummer model via progenitor stars that
greatly outmass their remnants.  This is one of the stiffest problems of
our formation scenario.

      If the dark cluster is less compact than a Plummer sphere, then its
half-mass radius must be smaller than 0.113 pc in order to fit the
kinematics.  All problems with stellar collisions get much worse.

      This problem also applies to all types of remnants. 

\subsubsection{White Dwarfs That Cannot Merge To Form Type Ia Supernovae
               Cannot Be Relevant}
 
      Interior to $r_h$, essentially all progenitors of 0.6 $M_\odot$
white dwarfs collide and either get destroyed (their surface escape
velocities are $\sim 600$ km s$^{-1}$), or, in the earlier phases of
the dark cluster formation, merge. If they merge, they get converted
to progenitors of white dwarfs that have masses \gapprox \ts0.8
$M_\odot$.  Progenitors of white dwarfs with masses \lapprox \ts0.55
$M_\odot$ live too long to have died and, provided they were not
destroyed, would still be visible.  Therefore, white dwarfs that are
low enough in mass so that a collision of two of them results in a
remnant that is less massive than the Chandrasekhar limit are not
relevant.

\subsubsection{Progenitors Collide And Evaporate or Merge Into High-Mass Stars}

      Figure 18 shows progenitor star collision times for the second
half of the dark cluster formation process.  At $r$ \lapprox
\ts$r_{1/4}$, progenitors of low-mass remnants get destroyed and
converted into gas, or, in earlier phases, merge to become progenitors
of high-mass remnants.  We note that the stellar evolution clock is
reset to essentially zero in every non-destructive collision, because the merging
stars get thoroughly mixed.  Dynamical friction is neglected in
constructing Figure 18; if it is included, then most of the
progenitors participate in the collisions.  Also neglected is the fact
that successive mergers increase the mass range and hence decrease
both the dynamical friction sinking time and the relaxation time of
the cluster.

     Three consequences spell trouble for the formation scenario:

     First, if progenitors are not destroyed, they merge up to form
stars of high enough masses so that they die as Type II supernovae.
Their luminosity is not a problem for the hypothetical, present
M{\ts}31 dark cluster, because its formation process is finished or at
least in hiatus.  But again, if M{\ts}31 is typical, then the
formation of many such objects at intermediate and high redshifts
should produce one Type II supernova per galaxy per $\sim$\ts$100$
years at the center of the galaxy.  They would have been seen.

     Second, the supernova ejecta again would not be bound to the dark
cluster unless a large amount of gas is also present.

     Third, relatively few, high-mass remnants would be formed.  Dynamical
friction would guarantee that they got deposited at small radii.  The
mass range that resulted from heterogeneous stellar merger histories would 
create a large mass range even if none was present intially.  The result 
would be that relaxation times would be much shorter -- plausibly an order
of magnitude shorter -- than the single-mass relaxation times that gave
rise to the cluster evaporation times in Figure 18.  For all of these
reasons, evaporation times are likely to be much shorter than the several
billion years indicated for 3 $M_\odot$ BHs in Figure 18.  This is 
implausibly short.

\subsubsection{Summary on BH alternatives}

     Therefore, astrophysically reasonable alternatives to a supermassive
black hole are likely to fail.  The arguments against brown dwarfs seem
bomb-proof.  The arguments against stellar remnants are more complicated, 
but they are based on secure aspects of stellar and star-cluster evolution.
Also, there are many arguments, even a few of which are sufficient.  
So we are not very vulnerable to uncertainties involving any one argument 
(``Are we sure that we have not missed those supernovae or confused them with
AGN activity?'').  The problem (\S\ts8.2.5) that remnants are deposited at
excessively small radii is particularly important.  In this paper, we have
derived the largest published estimate of $M_\bullet$ using data at the smallest
radii.  This leaves little room for distributed dark matter; i.{\ts}e., for a
dark cluster with core-halo structure. In addition, \S\ts8.2.3 on the formation
scenario, while not formally part of our argument against dark clusters, 
presents formidable challenges.  Our arguments are discussed in more detail 
in Kormendy et al.~(2005).  However, our conclusion that dark cluster
alternatives to a BH are excluded seems robust.

\section{Conclusion}

      M{\thinspace}31 is now the third galaxy in which astrophysical arguments
strongly favor the conclusion that a dynamically detected central dark object
is a BH.  M{\thinspace}31 is the only galaxy for which such arguments are based
on HST observations.  Similar conclusions for NGC 4258 and for our Galaxy 
result from ground-based observations.  The present result is therefore an
important contribution of HST to the BH paradigm of active galactic nuclei. 
It increases our confidence that all dynamically detected central dark objects
in galaxies are black holes.

\acknowledgments

      We are most grateful to Don Winget for very helpful discussions
about white dwarf properties and to Don and to Anjum Mukadam for
providing digital observed spectra of white dwarf stars. We thank
Detlev Koester for making his models of white dwarf atmospheres
available and for insightful remarks about their systematics.  We also
thank the Nuker team (D.~Richstone, PI) and especially Scott Tremaine
for many helpful and enlightening discussions. R.B. is most grateful
to the Department of Astronomy of the University of Texas at Austin
for its warm hospitality and the support provided by a Beatrice 
M.~Tinsley Centennial Visiting Professorship. We thank the anonymous
referee for a careful reading of the paper and for very helpful suggestions
that led to significant improvements in \S\S\ts1 and 8 and that 
prompted us to add \S\ts6.2.

This paper is based on observations made with the NASA/ESA Hubble
Space Telescope, obtained at the Space Telescope Science Institute,
which is operated by the Association of Universities for Research in
Astronomy, Inc., under NASA contract NAS5--26555. These observations
are associated with program 8018 (visits 1 and 2) and were a portion
of time allocated to the Space Telescope Imaging Spectrograph
Instrument Definition Team key project for supermassive black holes.


\vfill\eject\clearpage

\appendix

\vskip 10pt plus 2pt minus 1pt

\centerline{\small LOSVD EVIDENCE FOR TREMAINE'S MODEL OF THE DOUBLE NUCLEUS}
\centerline{\small AS AN ECCENTRIC DISK OF STARS ORBITING THE BH}

\vskip 10pt plus 2pt minus 1pt

      Peiris \& Tremaine (2003) show that their eccentric disk model
fitted to ground-based kinematic data also agrees remarkably well with
our STIS kinematic measurements of P1 and P2.  The comparison includes
not only $V$ and $\sigma$ but also the parameters $h_3$ and $h_4$
which measure the lowest-order departures from Gaussian line profiles.
The data that they use are presented here in Tables 1 and 2 
and in Figures 19 and 20.  We will not repeat
their discussion.  Instead, we focus on the generic properties of the
line-of-sight velocity distributions (LOSVDs).  In particular, we
confirm an unusual property of the LOSVDs that directly implies
aligned, eccentric orbits.  This effect was seen and interpreted in
KB, but it is much larger at HST spatial resolution.  Since the effect
was inherent in but not explicitly predicted by Tremaine (1995), it is
compelling evidence in favor of his model.



\begin{figure*}[b]
\centerline{\psfig{file=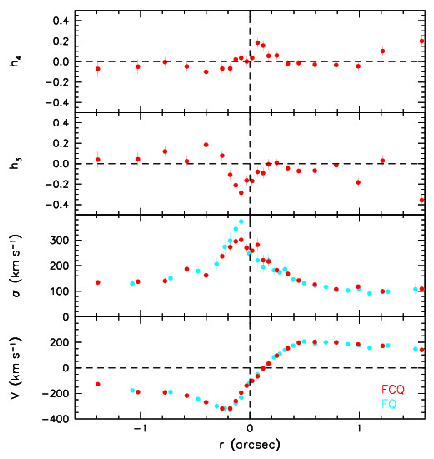,width=13.5cm,angle=0}}
\figcaption[m31_vsigh3h4_jk.cps]{The red points show the rotation velocity 
$V$, velocity dispersion $\sigma$, and Gauss-Hermite parameters $h_3$ and 
$h_4$ as a function of radius as derived with FCQ from the red spectrum of 
the nucleus of M{\thinspace}31.  The blue points are the FQ results from
Figure 2. 
\label{fig19}}
\vskip 0.3cm
\end{figure*}




\begin{figure*}[t]
\centerline{\psfig{file=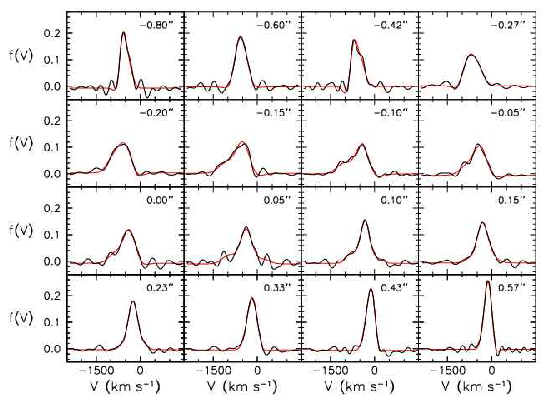,width=18.5cm,angle=-90}}
\figcaption[m31losvd.cps]{Line-of-sight velocity distributions in the same 
radial bins that were used to derive the FCQ kinematic results shown in 
Figure 19.  The black lines are nonparametric LOSVDs; the red curves are the
FCQ fits.  The radius of the bin is given at upper-right in each panel.
At upper-left, we tabulate the velocity dispersion and the Gauss-Hermite
parameters $h_3$ and $h_4$.
\label{fig18}}
\end{figure*}


      Figure 19 compares the FQ and FCQ reductions of the Ca infrared triplet
spectra.  The velocities are almost identical.  The dispersions agree where the
higher-order Gauss-Hermite coefficients are small and differ in the expected way
when they are not.  Where FCQ measures an extended wing of the
LOSVD in the prograde direction ($h_3 < 0$ at $r \simeq -0\farcs1$), it finds a
smaller dispersion than FQ, because FQ fits a Gaussian to the whole LOSVD,
including wings.  Similarly, where $h_4 > 0$ at $r \simeq +0\farcs1$, FCQ fits
a broader Gaussian than FQ and then clobbers the intermediate-velocity shoulders
of the Gaussian with $h_4$ to make the fitted LOSVD triangular.

      Kinematic asymmetries inherent in an eccentric disk provided the basic
test of that model in Tremaine (1995), in KB, and in Peiris \& Tremaine (2003).
We now know that the BH is in the blue nucleus P3 at $r = 0$ in Figure 19.  Stars
in the eccentric disk linger at apocenter to form P1; as a result, both the
rotation amplitude $|V|$ and the velocity dispersion $\sigma$ are relatively
small at $r \simeq +0\farcs5$.  Other stars in the eccentric disk are passing
pericenter slightly on the anti-P1 side of the BH; as a result, the rotation
amplitude is large at $r \simeq -0\farcs2$ in P2.  The apparent velocity
dispersion is highest at approximately the same radii because the slit and PSF
average over stars moving in a variety of directions as they swing around the
BH.  Figures 11 and 12 in Peiris \& Tremaine (2003) show that their nonaligned
model accurately accounts for the asymmetric rotation and dispersion profiles.

      An important additional property of the LOSVDs -- and the unusual one
mentioned above -- is the observation that $h_3$ has the same sign as $V$ over 
a radius range of $\Delta r \simeq 0\farcs$4 centered $\sim 0\farcs$1 on the
anti-P1 side of P3.  This effect was seen at ground-based spatial
resolution in KB. It is much larger here. It is opposite to the normal behavior
of rotating stellar systems, in which velocity projection along the line of
sight makes $h_3$ antisymmetric with $V$ (van der Marel et al.~1994; Bender,
Saglia, \& Gerhard 1994).  Also, the maximum amplitude, $h_3 \simeq -0.3$, is
unusually large compared to values in other galaxies.  All this is easily seen
in the LOSVDs (Figure 20; radii $-$0\farcs15 and $-$0\farcs10) as the broad
wings on the $-V$ side of the line centers.  These wings tell us that, on the
anti-P1 side of the P3, where the average galactic rotation is toward
us ($V$ is negative in Figure 20), many stars are rotating more rapidly and few
stars are rotating more slowly than the mean rotation velocity averaged within
the PSF.

      Our interpretation is the same as in KB.  The velocity
dispersion is expected to look big near pericenter in the eccentric
disk, because the slit and PSF integrate over stars that are at
different positions along orbits that are rapidly turning around the
BH.  A prograde LOSVD wing follows naturally if there are many stars
still closer to the BH and if they also are in eccentric orbits with
apocenters that point toward P1.  In the almost-Keplerian potential of
the BH, these stars have larger pericenter velocities than the mean
$V$ farther out; in fact, their velocities should be larger than the
local circular velocity.  Consistent with this interpretation, the
LOSVD asymmetry is most obvious at $r = -0\farcs05$ to $-0\farcs15$,
i.{\thinspace}e., at slightly more than one PSF radius on the anti-P1
side of P3. The highest velocities reach $\sim 1000$ km s$^{-1}$,
indeed somewhat larger than what we measure at about the same location
for the PSF-blurred velocities of the blue stars in P3. Of
course, this explanation only works if the BH is embedded in P3.

      The fact that we can understand naturally an observation not
predicted by Tremaine (1995) increases our confidence in his model.
With improved disk parameters, Peiris \& Tremaine (2003, see Figure~13
and 14) accurately predict the $h_3$ and $h_4$ profiles near the BH.
At this point, there seems little doubt that the interpretation of the
double nucleus as an eccentric disk is correct and that its main
parameters have been determined.  The important next step is
self-consistent dynamical models to investigate whether the present
configuration can be long-lived (e.{\ts}g., Statler et al.~1999;
Statler 1999; Bacon et al.~2001, Salow \& Statler 2004). A larger
black hole mass. as estimated here, will likely help to construct more
long lived models. Beyond that, the origin of the eccentric disk
remains essentially unknown.

\vskip 0.9cm

\begin{center}
\textsc{Table 4}

\vskip 0.1cm 

\textsc{Kinematics of M~31 derived from the red CaT spectra
with the Fourier Quotient method}

\vskip 0.2cm 

\begin{tabular}{rrrrr}
\tableline
\tableline
\tablevspace {3pt}
     radius   &    V~~  &   $\Delta$V~~ &  $\sigma$~~ & $\Delta\sigma$~~\\   
     arcsec   &   km s$^{-1}$ & km s$^{-1}$ & km s$^{-1}$ & km s$^{-1}$ \\ 
\tablevspace {2pt}
\tableline 
\tablevspace {3pt}
    $-1.075  $&$  -177  $&    14  &130   &15 \\
    $-0.726  $&$  -189  $&    13  &151   &14 \\
    $-0.474  $&$  -244  $&    13  &179   &14 \\
    $-0.304  $&$  -299  $&    23  &207   &24 \\
    $-0.230  $&$  -319  $&    32  &274   &35 \\
    $-0.180  $&$  -332  $&    34  &298   &38 \\
    $-0.130  $&$  -276  $&    36  &345   &39 \\
    $-0.080  $&$  -232  $&    45  &373   &48 \\
    $-0.005  $&$  -118  $&    23  &247   &25 \\
    $ 0.070  $&$   -52  $&    28  &221   &31 \\
    $ 0.120  $&$     0  $&    22  &195   &23 \\
    $ 0.170  $&$    32  $&    23  &220   &26 \\
    $ 0.220  $&$    84  $&    21  &182   &23 \\
    $ 0.270  $&$   108  $&    14  &174   &15 \\
    $ 0.320  $&$   140  $&    15  &186   &17 \\
    $ 0.396  $&$   175  $&    10  &146   &11 \\
    $ 0.495  $&$   207  $&     8  &131   &8  \\
    $ 0.595  $&$   192  $&     7  &122   &8  \\
    $ 0.694  $&$   200  $&     7  &117   &8  \\
    $ 0.794  $&$   200  $&     7  &104   &8  \\
    $ 0.894  $&$   185  $&     9  &103   &11 \\
    $ 0.994  $&$   178  $&    11  &111   &13 \\
    $ 1.094  $&$   156  $&    10  &91    &12 \\
    $ 1.262  $&$   175  $&    11  &99    &13 \\
    $ 1.513  $&$   147  $&    19  &108   &23 \\
\tablevspace {2pt}
\tableline
\end{tabular}
\end{center}

\vskip 0.8 cm

\begin{center}
\textsc{Table 5}

\vskip 0.1cm 

\textsc{Kinematics of M~31 derived from the red CaT spectra
with the Fourier Correlation Quotient method}

\vskip 0.2cm 

\begin{tabular}{rrrrrrrrr}
\tableline
\tableline
\tablevspace {3pt}
     radius   &    V~~  &   $\Delta$V~~ &  $\sigma$~~ & $\Delta\sigma$~~& $h_3$~~
     & $\Delta h_3$~~ & $h_4$~~ & $\Delta h_4$~~ \\   
     arcsec   &   km s$^{-1}$ & km s$^{-1}$ & km s$^{-1}$ & km s$^{-1}$ & & & & \\ 
\tablevspace {2pt}
\tableline 
\tablevspace {3pt}
  $-1.390$  &  $-130.8$  & 11.7  & 135.4  & 10.6  &  $ 0.040$  &  0.079  &   $-0.092$  &  0.079 \\
  $-1.023$  &  $-189.6$  & 11.0  & 140.0  & 11.6  &  $ 0.056$  &  0.072  &   $-0.043$  &  0.072 \\
  $-0.776$  &  $-192.2$  &  8.3  & 144.2  & 10.2  &  $ 0.110$  &  0.053  &   $ 0.013$  &  0.053 \\
  $-0.578$  &  $-217.0$  & 11.2  & 188.7  & 11.8  &  $ 0.037$  &  0.054  &   $-0.046$  &  0.054 \\
  $-0.401$  &  $-267.2$  &  7.2  & 168.3  &  7.1  &  $ 0.168$  &  0.039  &   $-0.070$  &  0.039 \\
  $-0.254$  &  $-316.4$  & 12.7  & 240.2  & 12.2  &  $ 0.082$  &  0.048  &   $-0.074$  &  0.048 \\
  $-0.180$  &  $-313.6$  & 14.4  & 279.6  & 15.5  &  $-0.089$  &  0.047  &   $-0.035$  &  0.047 \\
  $-0.130$  &  $-247.9$  & 16.3  & 296.5  & 22.1  &  $-0.188$  &  0.050  &   $ 0.060$  &  0.050 \\
  $-0.080$  &  $-181.4$  & 16.1  & 300.7  & 23.1  &  $-0.264$  &  0.049  &   $ 0.084$  &  0.049 \\
  $-0.030$  &  $-137.6$  & 14.5  & 273.7  & 18.2  &  $-0.147$  &  0.048  &   $ 0.026$  &  0.048 \\
   $0.020$  &   $-99.0$  & 13.0  & 263.9  & 17.3  &  $-0.156$  &  0.045  &   $ 0.051$  &  0.045 \\
   $0.070$  &   $-59.6$  & 15.6  & 286.1  & 27.5  &  $-0.061$  &  0.050  &   $ 0.194$  &  0.050 \\
   $0.120$  &    $-4.8$  & 12.6  & 226.0  & 21.5  &  $-0.072$  &  0.051  &   $ 0.176$  &  0.051 \\
   $0.170$  &    $36.0$  & 13.0  & 231.1  & 18.4  &  $-0.001$  &  0.051  &   $ 0.079$  &  0.051 \\
   $0.246$  &    $96.2$  &  8.4  & 184.9  & 11.5  &  $ 0.009$  &  0.041  &   $ 0.062$  &  0.041 \\
   $0.346$  &   $152.2$  &  7.0  & 170.3  &  8.2  &  $-0.042$  &  0.038  &   $-0.007$  &  0.038 \\
   $0.446$  &   $198.2$  &  4.8  & 143.3  &  5.7  &  $-0.075$  &  0.031  &   $-0.002$  &  0.031 \\
   $0.592$  &   $203.6$  &  3.5  & 126.1  &  3.9  &  $-0.072$  &  0.025  &   $-0.021$  &  0.025 \\
   $0.788$  &   $199.2$  &  3.5  & 109.2  &  3.8  &  $-0.016$  &  0.029  &   $-0.029$  &  0.029 \\
   $0.989$  &   $185.5$  &  5.7  & 117.0  &  6.1  &  $-0.185$  &  0.045  &   $-0.037$  &  0.045 \\
   $1.213$  &   $173.6$  &  5.9  & 103.1  &  8.2  &  $ 0.014$  &  0.052  &   $ 0.075$  &  0.052 \\
   $1.569$  &   $140.3$  &  7.0  & 108.7  & 12.5  &  $-0.379$  &  0.059  &   $ 0.221$  &  0.059 \\
\tablevspace {2pt}
\tableline
\end{tabular}
\end{center}

\end{document}